\documentclass[review]{elsarticle}

\usepackage{amssymb}
\usepackage{amsmath}
\usepackage{hyperref}
\usepackage{color}
\definecolor{comstumGreen}{rgb}{0.0, 0.50, 0.0}
\usepackage{float}
\usepackage{tikz}
\usepackage{graphicx}
\usepackage{subcaption}
\usepackage{pgfplots}
\usetikzlibrary{plotmarks}
\journal{Elsevier}

\begin{document}

\begin{frontmatter}

\title{Numerical simulation of two-phase slug flows in horizontal pipelines: A 3-D smoothed 
particle hydrodynamics application}

\author[mymainaddress]{Massoud Rezavand\corref{mycorrespondingauthor}}
\cortext[mycorrespondingauthor]{Corresponding author}
\ead{massoud.rezavand@tum.de}

\author[mymainaddress]{Xiangyu Hu}
\ead{xiangyu.hu@tum.de}

\address[mymainaddress]{Department of Engineering Physics and Computation, TUM School of Engineering and Design, Technical University of Munich, 85748 Garching, Germany}

\begin{abstract}
A fundamental difficulty of studying gas-liquid pipe flows is the prediction of the occurrence and characteristics 
of the slug flow regime, which plays a crucial role in the safety
design of oil pipelines. Current empirical methods and one-dimensional computational
models only achieve limited success. While
3-D numerical simulations are highly recommended, 
they have been very seldom used. We perform 3-D Lagrangian numerical simulations of gas-liquid pipe flows,
and focus on the interfacial instabilities leading to slug formation. We adapt an existing 
multi-phase smoothed particle hydrodynamics (SPH) method based on a Riemann solver to achieve an efficient solver
with no dependency on empirical correlations. To realize the 
high inlet velocities of gas and liquid an in- and outlet boundary condition is presented. 
The results are validated against existing experimental data, numerical simulations and analytical solutions.
Multiple gas-liquid pipe flow patterns are predicted, namely, smooth stratified, stratified wavy, 
bubble flow, slug flow and bubble flow. Several principle characteristics of slug flows, e.g., pressure gradient, slug development
and slug frequency are analyzed.

\end{abstract}


\end{frontmatter}


\section{Introduction}\label{sec:introduction}
Gas-liquid multi-phase flows in piping systems are ubiquitous in a wide range of industrial applications, e.g., 
nuclear power plants, crude oil mixed transportation, condensation, distillation towers 
and filling or emptying operations among others.
At high flow rates, this category of flows can often demonstrate flow characteristics that might play a destructive role 
in pipelines. For this reason it is of high importance to study and understand them well,
in order to prevent the undesired phenomena \cite{sausen2012slug,KirsnerProceedings}.

Given various physical conditions, such as 
flow rate, pipe diameter and inclination and mixture fractions, multi-phase pipe 
flows experience significantly different characteristics and flow regimes. 
One of the most common flow patterns that occurs in petroleum and chemical industries is 
the formation and migration of liquid slugs in piping systems. 
For instance, the flow conditions governing most the transportation processes
in oil wells lie within the range of slug flow regime \cite{bonizzi2003transient}.
A fundamental difficulty of studying such flows is on predicting the occurrence and characteristics
of the flow regimes \cite{Fabre-annurev.fl.24.010192.000321,ujang2003studies}. 
For gas-liquid pipe flows, in particular, such difficulty is on predicting when
the slug flow regime occurs and how it is characterized, which are highly relevant to the safety
design of the oil pipelines. Several flow regime prediction models have been developed to describe 
two-phase pipe flow patterns and flow transitions, which are either empirical or based on experimental observations
(e.g., \cite{mandhane1974flow,mishima1980theoretical,barnea1987unified,taitel1976model}). 
Despite being substantially useful and widely 
applied for designing industrial applications, these models are not adequate for a complete flow definition, as 
they do not represent the interaction of various forces involved in the flow. 

In another approach, numerous studies have been dedicated to develop and apply one-dimensional (1-D) 
thermal hydraulic system codes to represent flow patterns and transitions (e.g., \cite{athlete3-2,chung2010mars,bajorek2008trace,BloemelingNeuhausSchaffrath2013,LEE2022112066}).
1-D solvers are useful in predicting the overall behavior of the process systems, however, they consider
major simplifications and are thus not appropriate for comprehensive flow analysis. They generally  
assume a fully developed flow and neglect details of interfacial area, mass transfer, viscous energy dissipation 
and turbulence effects and treat the flow governing equations in a simplified manner. For this reason, 
some studies have attempted to couple such models with more comprehensive fluid flow solvers to fill the gap
(see e.g., \cite{ishii1975thermo,lu2015experimental,korzilius2017modeling}). 

Intermittent flows have also been extensively studied experimentally considering various aspects of the problem
(e.g., \cite{vallee2008experimental,ABDULKADIR2016147,wu2021_frontiers}). However, 
experimental techniques are costly, resource-intensive and not applicable for all full-scale piping systems. 
Thanks to modern high performance computers, computational fluid dynamics (CFD) methods have emerged as promising tools to study multi-phase pipe flows (see e.g., \cite{vallee2008experimental,wu2021_frontiers,ramdin2012computational}).
In order to simulate slug flow problems, the main challenge of the traditional Eulerian mesh-based methods is to
deal with its multi-phase nature and the interfacial complexities. For instance, the volume of fluid (VOF) approach has been 
implemented in commercial packages by Taha and Cui \cite{TAHA20041181} and Al-Hashimy et al. \cite{al2016numerical}
to simulate two-phase slug flows in capillaries and pipelines, respectively. The Level-set (LS) and lattice-Boltzmann (LB)
methods were also employed to study slug flows in micro-channels for the 
applications in microfluidic devices \cite{FUKAGATA200772,YU20077172}. 
Computationally more expensive methods, e.g., direct numerical simulations (DNS)
 have also been used for the simulation of 
two-phase slug flows in inclined pipes. Xie et al. \cite{xie_zheng_2017} also compared two-dimensional (2-D) results with three-dimensional (3-D) ones in their DNS studies and concluded that there are differences between 2-D
and 3-D simulations in terms of circulation dynamics and vorticity generation.

In recent decades, Lagrangian mesh-free methods have emerged as an attractive alternative for 
mesh-based methods in the simulation of multi-phase flows (see e.g., \cite{rezavand2018isph,Lou-Khayyer.2022.jc873,PATINONARINO2023104355}). 
As an advanced member of this class of methods, 
smoothed particle hydrodynamics (SPH) is a fully Lagrangian particle-based method that 
has been proposed in late 1970s for astrophysics. SPH has already been successfully used for a wide range of theoretical and 
engineering problems, ranging from free-surface flows to human cardiac function simulations (see e.g., 
\cite{REZAVAND2020109092,ZHANG_sphinxsys,ZHANG2021_cardiac,KHAYYER202384})
Owing to its Lagrangian nature, SPH naturally realizes the 
multi-phase interface with
no need for an additional algorithm. These peculiarities nominates SPH as a promising choice for 
simulating multi-phase pipe flows and has been already employed for slug flow simulations (e.g., \cite{korzilius2017modeling,DOUILLETGRELLIER2018101,douillet2019comparison}). 

Although the previous SPH simulations of intermittent pipe flows have 
shed light on the nature of slugging from a 
Lagrangian point of view, those studies are limited to 1-D or 2-D frameworks.
To the best of our knowledge, the 3-D Lagrangian numerical investigation of slug flows is still unaddressed.  
As Xie et al. \cite{xie_zheng_2017}
concluded, 3-D simulations of slug flows are of importance for a comprehensive 
representation of the flow structures.
On the other hand, as a well-established turbulence model has not yet 
been presented for the SPH formulation in
industrial scales, it is important to carry out 3-D SPH simulations in order to 
capture the large vortices, relying 
on the inherent large-eddy simulation (LES) turbulence mechanisms of SPH \cite{GhasemiV_2013}. 
Cleary and Monaghan \cite{cleary1993boundary} proposed that dissipation mechanisms
exist at the sub-particle scales of the SPH scheme that cause energy dissipation, thereby preventing 
excessive energy accumulation at the larger scales.
Ting et al. \cite{ting2005simulation} and Ghasemi et al. \cite{GhasemiV_2013} confirmed this proposal with their SPH 
simulations of turbulent flows and considered SPH as a natural LES method. Furthermore, 
they realized that for more accurate results 3-D computations should be conducted.  

The above considerations are the motivations of the 
present study to perform robust 3-D simulations of gas-liquid pipe flows within the fully Lagrangian 
framework of the SPH method. The application of SPH for such an industrial problem will also address
the fifth grand challenge of the SPH method \cite{vacondio2021grand}: applicability to industry. 
In addition, we aim at presenting a 
numerical framework with no empirical 
dependency with a relatively good computational efficiency to fill the gap highlighted by the comprehensive studies of 
Mohmmed et al. \cite{MOHMMED2021116611}. To achieve these goals, we implemented our methodologies in the 
open-source SPHinXsys library \cite{zhang2021sphinxsys} 
being available with accompanying information at \url{https://www.sphinxsys.org}.

We firstly present an in- outflow boundary condition implemented in 
the open-source SPHinXsys library \cite{ZHANG_sphinxsys}, which enables us to apply desired 
inlet velocity of both gaseous and liquid phases. Secondly, the accuracy and convergence of the proposed 
numerical framework is quantitatively verified in comparison with analytical solutions. We next 
validate the presented method against experimental observations in five different gas-liquid pipe flow regimes. 
The initiation of the first slug at the entrance of the pipe has also been addressed here.
Finally, various characteristics of a slug flow regime in a horizontal pipe is investigated, namely, 
pressure and velocity distribution throughout the pipe, development of a slug along the pipe and slug frequency.
The obtained results by the presented SPH method are further compared with the available data sets in the literature
and the method demonstrates good accuracy and robustness.
 
\section{Numerical method}\label{sec:NumericalModel}
Intermittent two-phase flows, e.g., slug flow, are typically considered as
two-phase flows with large density ratio,
i.e., $ \rho_l/\rho_g \gg 1$, where $\rho_l$ and $\rho_g$ 
represent the densities of liquid and gaseous phases. 
Such problems have been investigated in our previous works in an inviscid regime 
\cite{REZAVAND2020109092, REZAVAND2022108507}. With the consideration of the 
viscous effects for slug flows,      
the mass and momentum conservation equations can be 
written respectively as
\begin{equation}\label{eq:massGeneral}
\frac{\text{d}\rho}{\text{d}t} = -\rho\nabla\cdot\mathbf{v},
\end{equation}
\begin{equation} \label{eq:momentumGeneral}
\frac{\text{d}\mathbf{v}}{\text{d}t} = -\frac{1}{\rho}\nabla p + \nu \nabla^2  \mathbf{v}+ \mathbf{g},
\end{equation}
where $\text{d}/\text{d}t$ is the Lagrangian derivative, $\rho$ the density, 
$\mathbf{v}$ the velocity, $p$ the pressure, $\nu$ the kinematic viscosity and $\mathbf{g}$ 
the gravitational acceleration. 
To close the system in a weakly compressible SPH (WCSPH) framework, pressure is estimated from density via an artificial equation of state (EoS), 
within the weakly compressible framework. Here, 
we use a simple linear equation for both the heavy and light phases
\begin{equation}\label{eq:EoS}
p = c^{2}(\rho-\rho_0),
\end{equation} 
where $c$ and $\rho_0$ are the artificial speed of sound and the initial reference density. 
Here, we assume that the speed of sound is constant 
and we set $c=10U_{max}$, where $U_{max}$ denotes the 
maximum anticipated velocity of the flow. Following the methodology
presented in Ref. \cite{REZAVAND2020109092} we use the same value of $c$ 
for both heavy and light phases, which leads to larger time-step sizes and thereby higher 
computational efficiency for the resource-intensive 3-D computations of the simulations 
presented in this work.

In order to cope with the complex two-phase phenomena in intermittent flows
we adapt the method presented in Ref. \cite{REZAVAND2020109092,REZAVAND2022108507}.
The particle summation formulation \cite{Monaghan_2012_Annual_Rew} is used 
to realize the mass conservation principle for the gaseous phase
\begin{equation}\label{eq:densitySummation}
\rho_i = m_i\sum_{j}W_{ij},
\end{equation}
where $m_i$ denotes the mass of particle $i$ and the smoothing kernel function 
$ W(\left| \mathbf{r}_{ij}\right| ,h)$ is simply substituted by $ W_{ij} $, 
with $\mathbf{r}_{ij}=\mathbf{r}_{i}-\mathbf{r}_{j}$ being the displacement vector between particle $i$ and $j$ 
and $h$ the smoothing length, 
respectively. 
The discretized form of the continuity equation is, on the other hand side, employed for the liquid phase
\begin{equation} \label{eq:SPH_massConservation}
\frac{\text{d}\rho_{i}}{\text{d}t} = \rho_i \sum_{j}\frac{m_j}{\rho_j}\mathbf{v}_{ij}\cdot\nabla_{i} 
W_{ij}=2\rho_i\sum_{j} \frac{m_j}{\rho_j} (\mathbf{v}_{i}- \mathbf{\overline{v}}_{ij}) 
\cdot\nabla_{i} W_{ij},
\end{equation}
where $\mathbf{v}_{ij}=\mathbf{v}_{i}-\mathbf{v}_{j}$ is the relative velocity 
and $\mathbf{\overline{v}}_{ij}=\frac{\mathbf{v}_{i}+\mathbf{v}_{j}}{2}$ denotes
the average velocity between particle $i$ and $j$.
Furthermore, the momentum conservation principle has been realized by the formulation presented in 
Ref. \cite{Monaghan_2012_Annual_Rew} excluding the artificial viscosity, 
\begin{equation} \label{eq:SPH_momentumConservation}
\frac{\text{d}\mathbf{v}_{i}}{\text{d}t} = -2\sum_{j} m_j \frac{\overline{{p}}_{ij}}{\rho_i\rho_j} \nabla_{i} W_{ij} +
2\sum_{j} \nu_i V_j \frac{\mathbf{v}_{ij}}{r_{ij}} \frac{\partial W_{ij}}{\partial r_{ij}} + \mathbf{g}
\end{equation}  
with $\overline{{p}}_{ij}=\frac{p_{i}+p_{j}}{2}$ being the average pressure between 
particle $i$ and $j$, $V$ the particle volume and $r_{ij}=\left| \mathbf{r}_i - \mathbf{r}_j \right|$. 

\subsection{WCSPH multi-phase Riemann solver for slug flow}
\label{subsec:multiphase_RiemannbasedSPH}
In WCSPH methods based on a Riemann solver \cite{vila1999particle,zhang2017weakly, REZAVAND2020109092}, 
an inter-particle Riemann problem is constructed 
along the unit vector $\mathbf{e}_{ij}=-\frac{\mathbf{r}_{ij}}{\left| \mathbf{r}_{ij}\right|}$ 
between the interacting particles $i$ and $j$ by piecewise constant approximation as follows
\begin{equation}\label{eq:LRstates}
\begin{cases}
(\rho_L, U_L, p_L) = (\rho_i, \mathbf{v}_i \cdot \mathbf{e}_{ij}, p_i)\\
(\rho_R, U_R, p_R) = (\rho_j, \mathbf{v}_j \cdot \mathbf{e}_{ij}, p_j)
\end{cases},
\end{equation}
The solution of the Riemann problem can be shown using a $x-t$ diagram, for which the readers are
referred to Ref. \cite{REZAVAND2020109092} to see more details.
For multi-phase flows with the same speed of sound across the interface, 
following Ref. \cite{hu2004interface}, the intermediate velocity and  pressure can be approximated as 
\begin{equation}\label{eq:RSsolution}
\begin{cases}
U^* = \overline{U} +\frac{p_L - p_R}{c(\rho_L+\rho_R)},\\
p^* = \overline{P} + \frac{\rho_L\rho_R\beta(U_L - U_R)}{\rho_L+\rho_R},
\end{cases}
\end{equation}
where $\overline{U} = (\rho_L U_L + \rho_R U_R)/(\rho_L+\rho_R)$ 
and $\overline{P} = (\rho_L p_R + \rho_R p_L)/(\rho_L+\rho_R)$.
   
For two-phase particle interactions with large density ratio, as in slug flow problems,
Eq. (\ref{eq:RSsolution}) gives approximately 
the intermediate velocity from the heavy phase, 
which indicates that the light phase experiences the heavy phase as a moving wall boundary, 
and intermediate pressure from the light phase, 
which indicates that the heavy phase undergoes a free-surface-like 
flow with variable free-surface pressure \cite{hu2004interface}.     
Having the intermediate values determined from the solution of the Riemann problem, 
the mass and  momentum conservation equations, 
i.e., Eqs. \eqref{eq:SPH_massConservation} and \eqref{eq:SPH_momentumConservation}, 
can be rewritten as
\begin{equation} \label{eq:SPH_star_massConservation}
\frac{\text{d}\rho_{i}}{\text{d}t} = 2\rho_i\sum_{j} \frac{m_j}{\rho_j} (\mathbf{v}_{i}- \mathbf{v}^{*}) \cdot\nabla_{i} W_{ij},
\end{equation}
\begin{equation} \label{eq:SPH_star_momentumConservation}
\frac{\text{d}\mathbf{v}_{i}}{\text{d}t} = -2 \sum_{j} m_j \frac{{p}^{*}}{\rho_i \rho_j} \nabla_{i} W_{ij} +
2\sum_{j} \nu_i V_j \frac{\mathbf{v}_{ij}}{r_{ij}} \frac{\partial W_{ij}}{\partial r_{ij}} + \mathbf{g}
\end{equation}
where $\mathbf{v}^{*} = U^{*} \mathbf{e}_{ij} 
+ (\overline{\mathbf{v}}_{ij} - \overline{U}\mathbf{e}_{ij} )$ and $\overline{\mathbf{v}}_{ij} 
= (\rho_i \mathbf{v}_i + \rho_j \mathbf{v}_j)/(\rho_i+\rho_j)$,
which is a density-weighted average velocity between particle $i$ and $j$.

\subsection{Gas-liquid pipe flow considerations}
Given gas and liquid phases, pipe geometry and inclination angle, the 
flow is determined by the superficial velocities $v_{sg}$ and $v_{sl}$. These properties are given by the gas and liquid 
volumetric flow rates $Q_g$ and $Q_l$, respectively
\begin{equation} \label{eq:slugRates}
v_{sg} =\frac{Q_g}{A}=v_{g}\alpha_g,~~~v_{sl} =\frac{Q_l}{A}=v_{l}\alpha_l,~~~\alpha_g+\alpha_l=1
\end{equation}  
where $v_{g}$ and $v_{l}$ are the gas and liquid velocities, respectively, $A$ is the cross-section area of
the pipe and $\alpha_g$ and $\alpha_l$ are the volumetric gas fraction and liquid fraction, or holdup, respectively.
The mixture velocity is then defined by the gas and liquid superficial velocities as 
\begin{equation} \label{eq:slugMixtureVel}
v_{m} =\frac{Q}{A}=v_{sg}+v_{sl}
\end{equation}  
where $Q$ is the overall flow rate.

\subsection{Boundary conditions}
\subsubsection{Solid wall}
In the present study, we use fixed dummy particles to impose the solid wall 
condition as proposed in Ref. \cite{REZAVAND2020109092} and further generalized for 
multi-phase flows in Ref. \cite{REZAVAND2022108507}. 
To realize the fluid-wall interactions 
a Riemann problem is constructed between particles of fluids and wall dummy particles, 
as for fluid-fluid interactions (see Section \ref{subsec:multiphase_RiemannbasedSPH}). 
However, the intermediate pressure value is obtained as
\begin{equation}\label{eq:RSatWall}
p^* = \frac{\rho_f p_w + \rho_w p_f}{\rho_f+\rho_w},
\end{equation}
where subscripts $f$ and $w$ denote fluid and wall, respectively, 
to decrease the wall-induced numerical dissipation while 
the intermediate velocity value $U^*$ is still obtained via Eq. \eqref{eq:RSsolution}, 
as for fluid-fluid interactions. 
Similar to Ref. \cite{ADAMI2012wall}, the pressure of 
wall dummy particles is calculated by the summation over all contributions of the 
neighboring fluid particles as
\begin{equation}\label{eq:wallPressure}
p_w = \frac{\sum_{f}\frac{p_f}{\rho_f} W_{wf} + (\mathbf{g}-\mathbf{a}_{w})\sum_{f} \mathbf{r}_{wf} 
	W_{wf}}{\sum_{f}\frac{W_{wf}}{\rho_f}}.
\end{equation}
where $\mathbf{a}_{w}$ denotes the wall acceleration.
It is worth noting that by introducing $\rho_f$ into Eq. \eqref{eq:wallPressure}
the contribution of liquid particles in $p_w$ at a triple point, 
where liquid, gas and solid particles meet, is vanishing.
As also presented by Adami et al. \cite{ADAMI2012wall}, 
the density of wall dummy particles is obtained from pressure via Eq. (\ref{eq:EoS}).

\subsubsection{In- and outflow}
\begin{figure}
	\centering
	\includegraphics[width=0.75\linewidth]{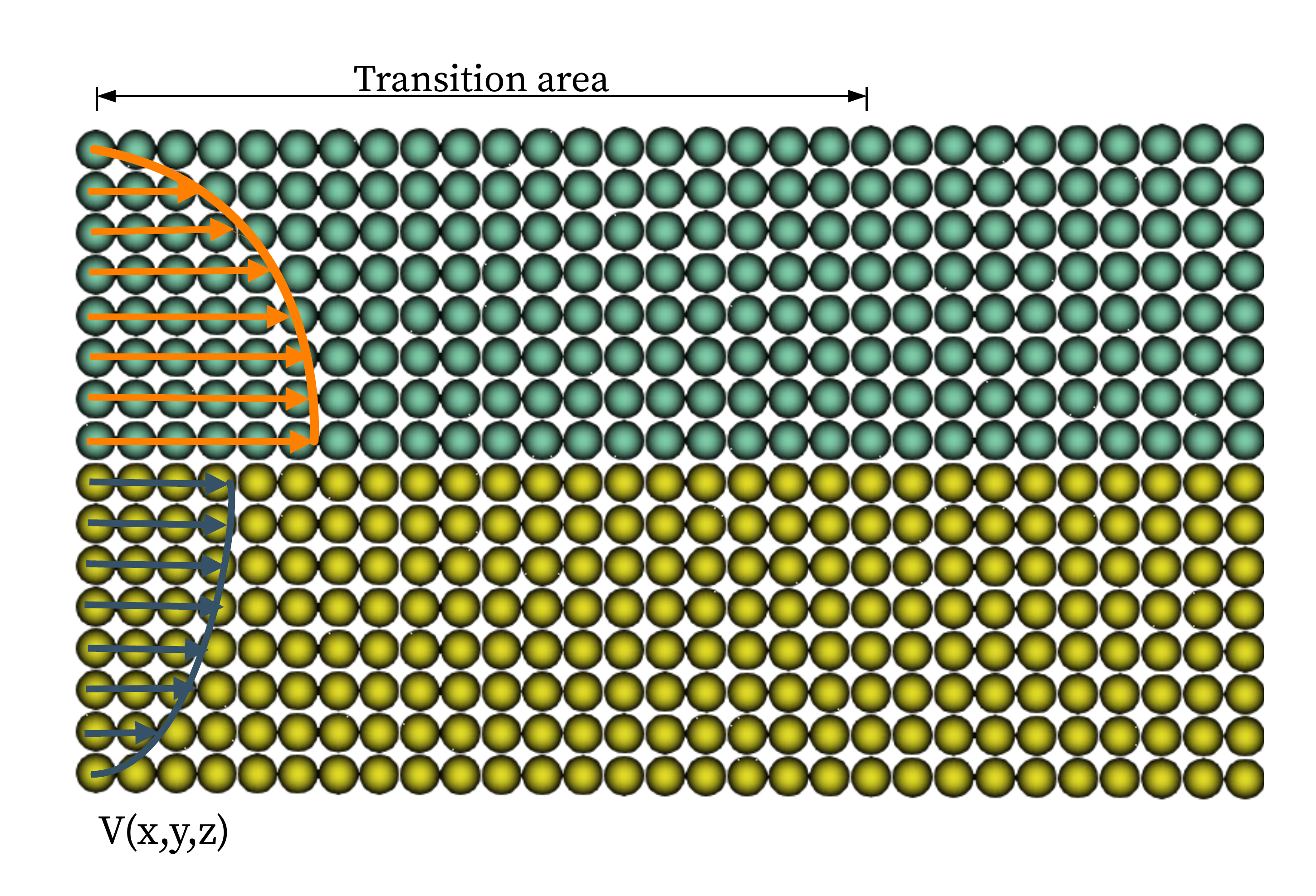}
	\caption{Schematic illustration of the in- and outlet boundary condition depicting different inlet velocity
		profiles for the gaseous (top) and liquid (bottom) phases together with the transition area. No-slip
	boundary condition is implemented on the surrounding solid walls.}
	\label{fig:inlet-outlet-schematic}
\end{figure}

The in- and outflow boundary condition using the emitter technique offered by SPHinXsys has been discussed 
in detail by Zhang et al. \cite{Shuoguo_arXiv}. In this section, we present another possibility in the library
to implement this boundary condition using a transition area. As shown in Fig. \ref{fig:inlet-outlet-schematic}, 
SPHinXsys considers a transition area, where the desired inlet velocity profile is applied on the particles.
This implementation takes advantage of the cell-linked list constructed for neighbor search procedure, 
where the inlet and outlet areas, located within $2h$ from the domain edge, are marked as cells 
where the particles need to be generated (at inlet) or removed (at outlet). In this manner, a periodic
boundary condition is realized, which is essential for an in- and outflow condition. 
The appropriate physical variables for particles in the inlet and outlet areas are then  
analytically applied and next extrapolated from within the flow domain, considering the smoothing kernel support 
radius \cite{Lastiwka2009permeable}.
As the inlet velocity magnitude in the test cases presented herein are relatively high, to have a stable inlet area
we define a minimum transition length of $20 dx$, where $dx$ is the initial particle spacing.
The configuration of the transition area together with the periodic boundary condition inlet and outlet areas
are then updated at every time step as discussed in Section \ref{sec:timeintegration}.

\subsection{Time integration}
\label{sec:timeintegration}
The time integration of the equations of motion in the present study employs the kick-drift-kick 
\cite{monaghan2005,adami2013transport} scheme. 
The first half-step velocity is obtained as
\begin{equation}
\mathbf{v}_i^{n + \frac{1}{2}} = \mathbf{v}_i^n +  \frac{\delta t}{2}  \big( \frac{\text{d} \mathbf{v}_i}{\text{d}t} \big)^{n},
\end{equation}
from which we obtain the position of particles at the next time step
\begin{equation}
\mathbf{r}_i^{n + 1} = \mathbf{r}_i^n +  \delta t  \mathbf{v}_i^{n + \frac{1}{2}}.
\end{equation}
With the updated flow states, the density of heavy-phase particles at the new 
time step is then calculated as
\begin{equation}\label{eq:newRho}
\rho_i^{n + 1} = \rho_i^n +  \delta t  \big( \frac{\text{d} \rho_i}{\text{d}t} \big)^{n+\frac{1}{2}},
\end{equation} 
where the time increment of density is approximated via Eq. \eqref{eq:SPH_massConservation}. 
It is worth noting that the density of light phase particles at the new time step 
is obtained using the density summation equation, i.e. Eq. \eqref{eq:densitySummation}. 
With the new densities being calculated, 
the pressure of particles and the time increment of velocity, can be obtained. 
As a final step to calculate the physical properties of the inner particles, 
the velocity of particles is updated for the new time step as
\begin{equation}
\mathbf{v}_i^{n + 1} = \mathbf{v}_i^{n + \frac{1}{2}} +  \frac{\delta t}{2}  
\big( \frac{\text{d} \mathbf{v}_i}{\text{d}t} \big)^{n+1}.
\end{equation}

In order to realize the in- and outlet boundary condition, the particle configuration at the inlet and outlet areas
will be updated according to the marked cells and the parabolic inlet velocity profiles will be imposed.
The analytical properties will be next extrapolated from the values within the flow domain.
To guarantee the numerical stability, the time-step size is limited by the CFL condition
\begin{equation}
\delta t \leq 0.25 \big( \frac{h}{c + U_{max}}  \big),
\end{equation}
together with the body force and viscous conditions
\begin{equation}
\delta t \leq 0.25 ~min\left( \sqrt \frac{h}{\left|\mathbf{g} \right| },\frac{h^2}{\nu } \right) .
\end{equation}

\section{Model validation}
\label{sec:modelValidation}
As the accuracy and robustness of the SPHinXsys library have been already demonstrated in various 
applications including fluid and solid dynamics \cite{ZHANG2020109135, dongWu2022, ZHANG_sphinxsys}, 
fluid structure interaction (FSI) \cite{ZHANG2021_multiFSI,REN2023113110} and biomechanics \cite{ZHANG20221_damping,ZHANG2021_cardiac},
in this section we consider two additional test cases, which showcase the ability of the library to 
capture the underlying physics of intermittent two-phase pipe flows. Firstly, we quantitatively verify the flow properties
in the Hagen-Poiseuille problem as a classical 3-D internal viscous flow. Secondly, 
the SPH simulations of a two-phase flow of gas and water under different velocity conditions in a
horizontal pipe are validated against experimental results carried out by Wu et al. \cite{wu2021_frontiers}.

\subsection{Hagen-Poiseuille flow}
\label{subsec:hagenPoiseuille}
\begin{figure}
	\centering
	\includegraphics[width=0.75\linewidth]{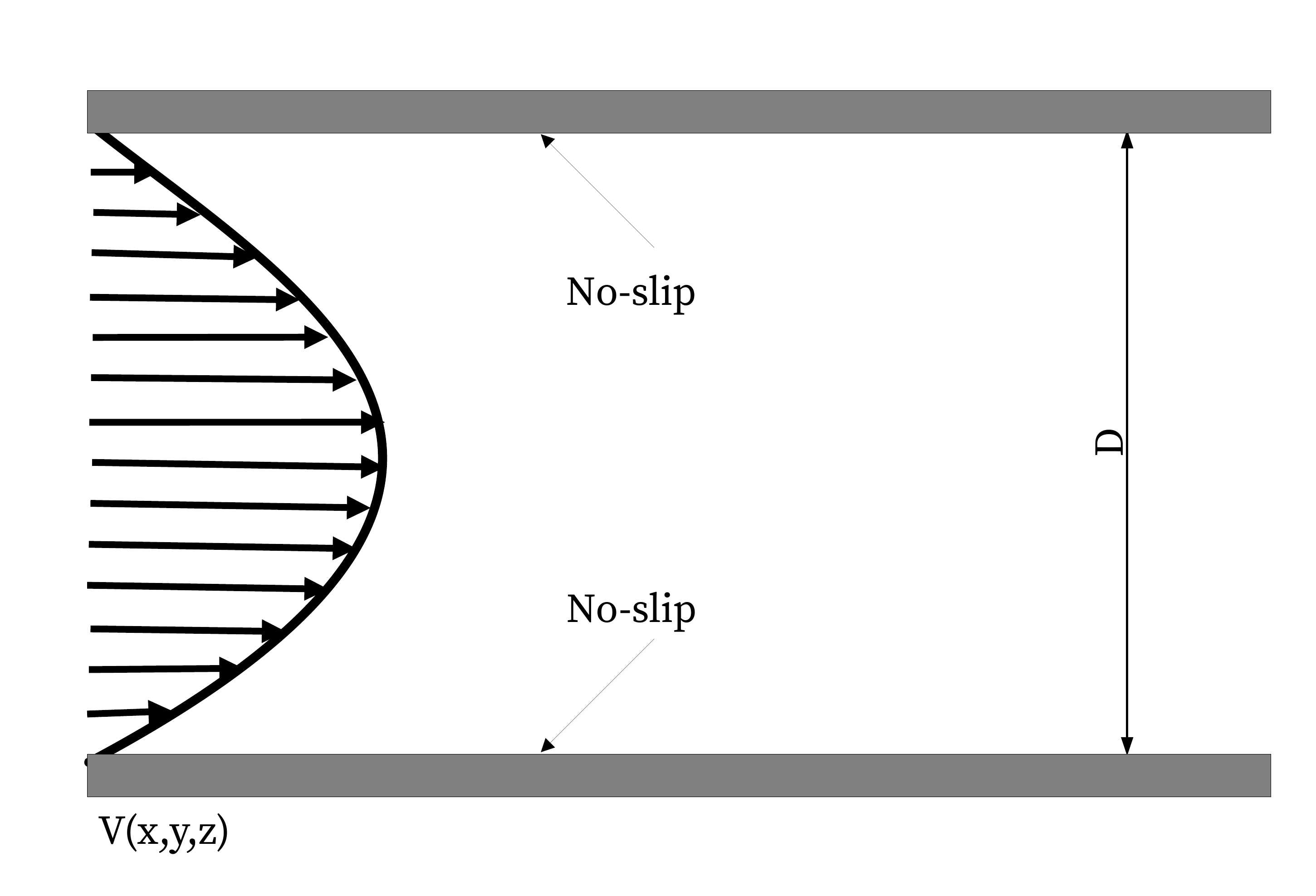}
	\caption{Schematic illustration of the 3-D Hagen-Poiseuille flow test case.}
	\label{fig:hagenPoiseuille_schematic}
\end{figure}

Due to the shear phenomena in Poiseuille flows both in 2- and 3-D, these problems are proper test cases to verify 
the viscosity model employed in the present methodology against analytical solutions \cite{MORRIS1997214, rezavand2018isph}. 
The schematic of this problem is illustrated in Fig. \ref{fig:hagenPoiseuille_schematic}. An incompressible liquid 
with a density of $\rho=1000~\mathrm{kg.m^{-3}}$ is flowing in the $x$-direction of a cylindrical channel with a diameter
of $d=1~\mathrm{mm}$ and the flow is driven by a horizontal free-stream velocity of $U_{max} = 0.1~\mathrm{mm.s^{-1}}$. 
The dynamic viscosity of the liquid is $\mu=0.001~\mathrm{mPa.s}$, which along with other physical properties 
determines the Reynolds number of the flow as $Re=100$. 
In this section, the periodic and no-slip boundary conditions are used for wall and horizontal boundaries, respectively. 

\begin{figure}
	\centering
	\begin{subfigure}[c]{0.45\textwidth}
		\centering
		\includegraphics[width=\linewidth]{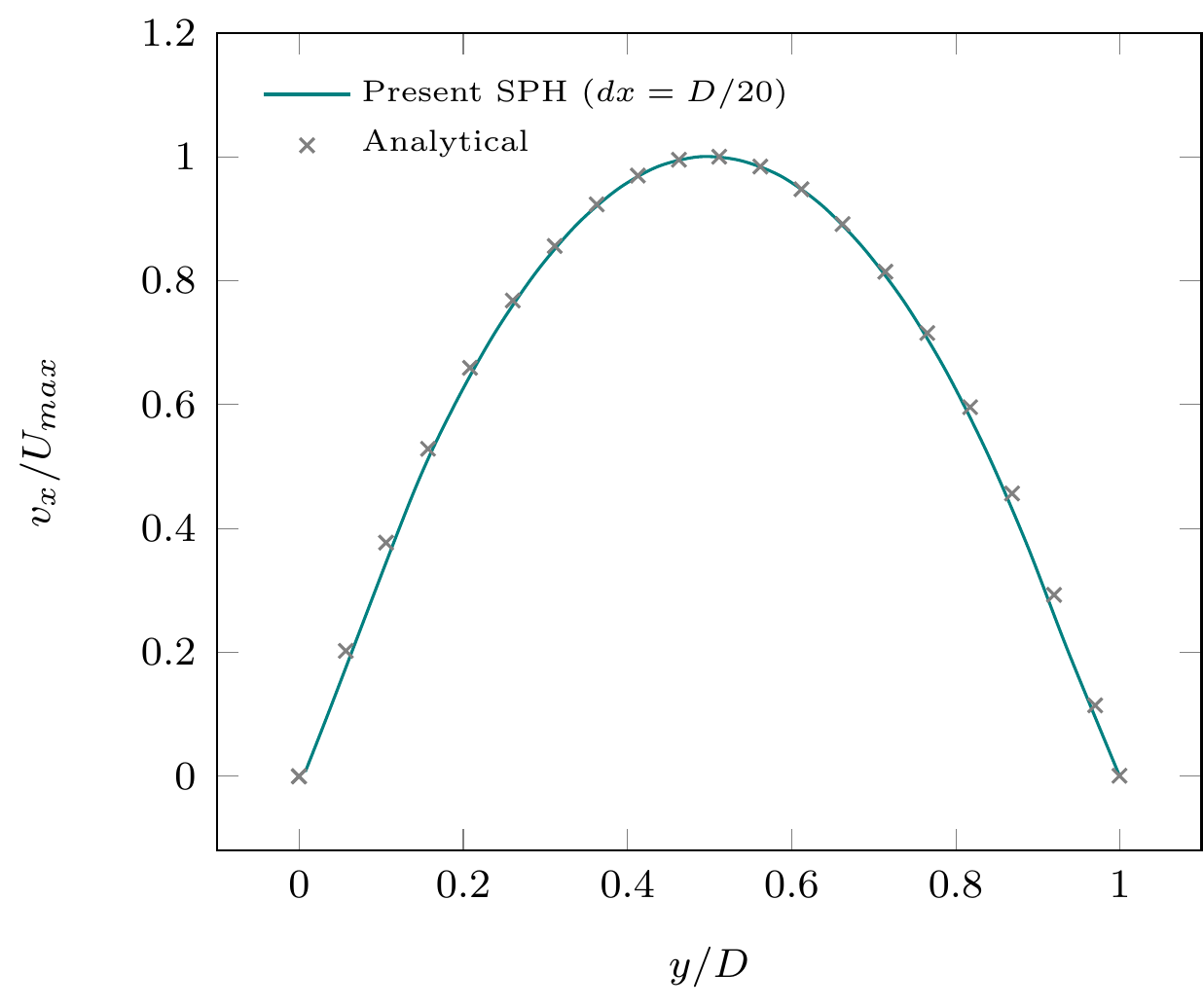}
		\caption{}
		\label{fig:hagenPoiseuille_vx_L2norm1}
	\end{subfigure}
	\hfill
	\begin{subfigure}[c]{0.5\textwidth}
		\centering
		\includegraphics[width=\linewidth]{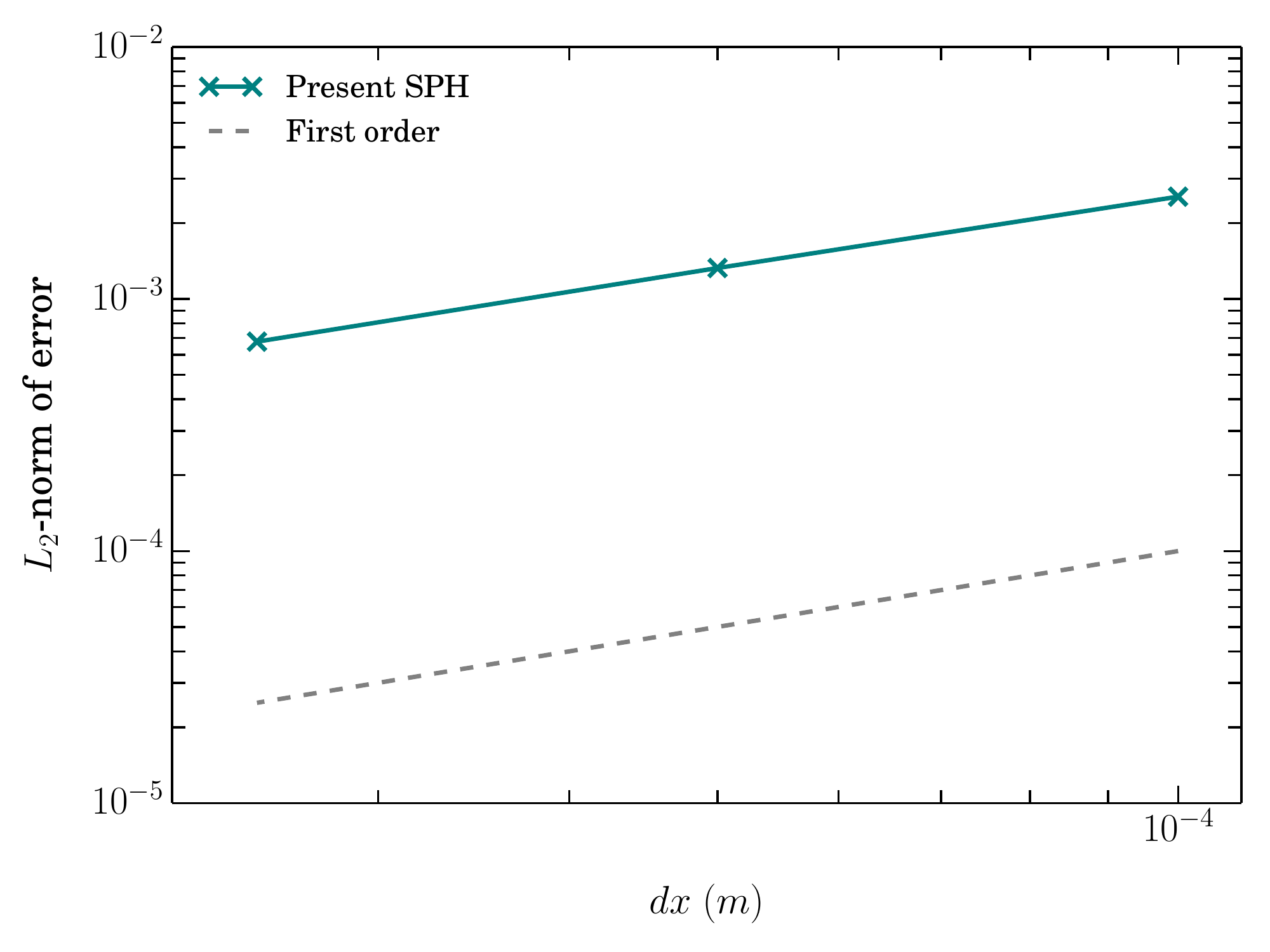}
		\caption{}
		\label{fig:hagenPoiseuille_vx_L2norm2}
	\end{subfigure}
	\caption{Hagen-Poiseuille flow: (a) Velocity profile obtained by the present SPH method in
		comparison with the analytical solution and (b) $L_2$-norm convergence analysis at $t=10~s$.}
	\label{fig:hagenPoiseuille_vx_L2norm}
\end{figure}

The velocity profile obtained from the single-phase SPH formulation presented in Section \ref{sec:NumericalModel} 
are compared with the analytical solutions derived by Bird et al. \cite{Bird_2002_transport} in Fig. \ref{fig:hagenPoiseuille_vx_L2norm1}. 
As shown in the figure, the SPH simulation with an initial particle spacing of $dx=D/20$ demonstrates 
a close agreement with the analytical solution. To further evaluate the numerical aspects of the 
proposed method, Fig. \ref{fig:hagenPoiseuille_vx_L2norm2} analyzes the convergence properties 
by the variation of the $L_2$-norm of error \cite{Fatehi_2011_error} with respect to the particles resolution. 
As it can be observed, 
an approximately first-order convergence is obtained.

\begin{figure}
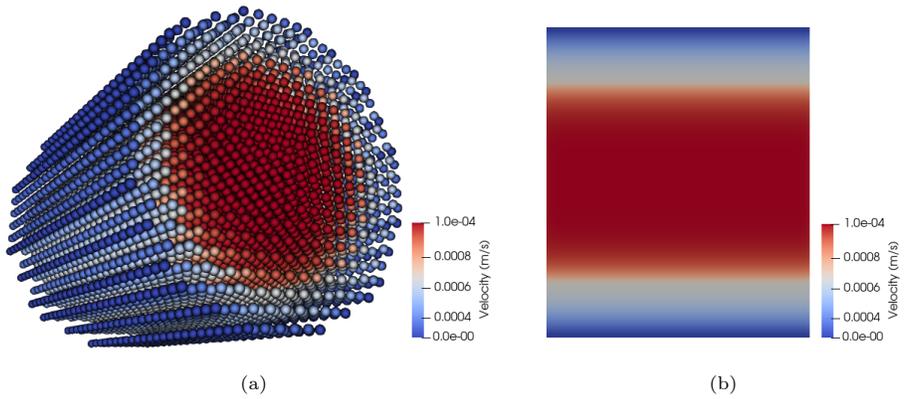

	\centering
	\begin{subfigure}[b]{0.55\textwidth}
		\centering
		\includegraphics[width=\linewidth]{Figs/Hagen_velocityProfile_1_3D.png}
		\caption{}
		\label{fig:hagenPoiseuille_3D_planeVelProf_a}
	\end{subfigure}
	\hfill
	\begin{subfigure}[b]{0.42\textwidth}
		\centering
		\includegraphics[width=\linewidth]{Figs/Hagen_velocityProfile_1.png}
		\caption{}
		\label{fig:hagenPoiseuille_3D_planeVelProf_b}
	\end{subfigure}
	\caption{Hagen-Poiseuille flow: (a) 3-D SPH particle distribution with particles being scaled by their velocity magnitude
			and (b) a $y$-normal cross section of the velocity profile at the same time instant with $dx=D/20$.}
	\label{fig:hagenPoiseuille_3D_planeVelProf}
\end{figure}

In order to investigate the viscous flow characteristics in more detail, 3-D SPH particle distribution 
and contours of the velocity magnitude with $dx=D/20$ are presented in 
Figs. \ref{fig:hagenPoiseuille_3D_planeVelProf_a} and \ref{fig:hagenPoiseuille_3D_planeVelProf_b}, respectively.
The particle distribution and the velocity field both demonstrate a smooth behavior with no spurious quantities.
Furthermore, the near wall velocity magnitude is noise-free, which demonstrates an accurate realization of 
the mass and momentum conservation equations and robust boundary condition implementations.

\subsection{Gas-liquid multi-phase flow in a pipe under various velocity conditions}
\label{subsec:twophasePipe}
\begin{figure}
	\centering
	\includegraphics[width=0.75\linewidth]{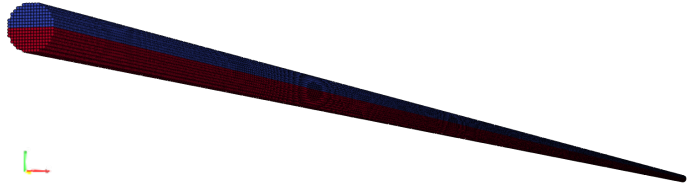}
	\caption{Gas-liquid multi-phase flow in a horizontal pipe: Initial geometric configuration of the 3-D horizontal pipe. 
		The upper half of the particles belong to the gaseous
		phase and the lower ones belong to the liquid phase.}
	\label{fig:slugInirialConfiguration}
\end{figure}

\begin{figure}
	\centering
	\includegraphics[width=1.\linewidth]{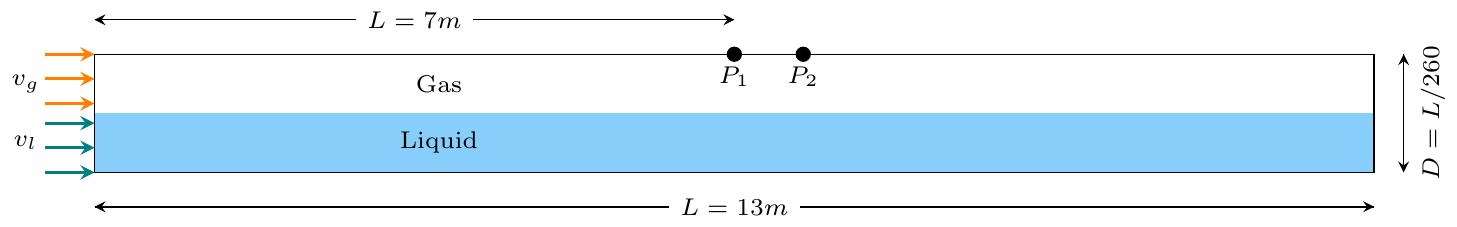}
	\caption{Gas-liquid multi-phase flow in a horizontal pipe: Schematic illustration of the problem. $P_1$ and 
		$P_2$ show the location of the two sensors used for recording pressure signals.}
	\label{fig:slug_schematic}
\end{figure}

To validate the presented SPH method against experiments, we used the
results obtained by Wu et al.\cite{wu2021_frontiers}. The experimental apparatus is described in their work with all details.
The apparatus uses a horizontal cylindrical tube with an inner
diameter of $50~mm$ and a length of $13~m$ ($260D$). Gas (air) and liquid (water) enter the
tube with constant flow rates through separated inlet cross-sections and mix by a stratified mixer with a 
length of $0.5~m$ at the inlet of the tube. The initial configuration of our SPH simulations is depicted in 
Fig. \ref{fig:slugInirialConfiguration} and the physical properties of the gaseous and the liquid phases are summarized 
in Table \ref{table:pipeflowProperties}. In order to have a more accurate understanding of the problem configuration,
the schematic view of the problems is also shown in Fig. \ref{fig:slug_schematic}, where the position 
of the pressure sensors used in the next sections are clarified, as well. 

\begin{table}
	\centering
	\begin{tabular}{l l l }
		\hline
		Physical property & Gaseous phase & Liquid phase  \\
		\hline
		Density $\mathrm{(kg.m^{-3})}$ & $1.204$ & $998.2$   \\
		Dynamic viscosity $\mathrm{(Pa.s)}$ & $1.837\times10^{-5}$ & $8.891\times10^{-4}$  \\
		Gravity $\mathrm{(m.s^{-2})}$& 9.81 & $9.81$  \\
		\hline
	\end{tabular}
	\caption{Physical properties of the light and dense fluids involved in the studied gas-liquid multi-phase 
		flow cases in horizontal pipe and under various velocity conditions.}
	\label{table:pipeflowProperties}
\end{table}

Depending on the superficial velocity conditions, various flow patterns will develop for which the flow characteristics are 
summarized in Table \ref{table:pipeflowCharacteristics}. In this section, we consider five different flow patterns
according to the Mandhane flow pattern diagram \cite{mandhane1974flow} depicted by Wu et al., namely, 
smooth stratified (SS), stratified wavy (SW), bubble flow (BF1 and BF2) and slug flow (SF).
\begin{table}
	\centering
	\begin{tabular}{l l c c}
		\hline
		ID & Flow pattern & $v_{sg}~(\mathrm{m.s^{-1}})$ & $v_{sl}~(\mathrm{m.s^{-1}})$  \\
		\hline
		SS & Smooth stratified & $0.5$ & $0.1$  \\
		SW & Stratified wavy & $2.5$ & $0.1$  \\
		BF1 & Bubble flow & $0.1$ & $3.0$  \\
		SF & Slug flow & $5.0$ & $1.0$  \\
		BF2 & Bubble flow & $0.3$ & $1.0$  \\
		\hline
	\end{tabular}
	\caption{Velocity condition and flow pattern description of the test cases presented in Fig. \ref{fig:5pipeflowPatterns}}
	\label{table:pipeflowCharacteristics}
\end{table}

\begin{figure}
	\scriptsize
	\centering
	\begin{tabular}{c c c c}
		ID &SPH & Experiment & OpenFOAM \\
		 SS & 
		 \includegraphics[width=0.3\linewidth]{Figs/intermittent_flow_patterns/SmoothStratified.png} & \includegraphics[width=0.3\linewidth]{Figs/intermittent_flow_patterns/SmoothStratified_exp.png} & \includegraphics[width=0.3\linewidth]{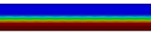} \\
		 SW & 
		 \includegraphics[width=0.3\linewidth]{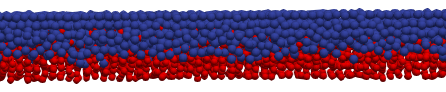} & \includegraphics[width=0.3\linewidth]{Figs/intermittent_flow_patterns/StratifiedWavy_exp.png} & \includegraphics[width=0.3\linewidth]{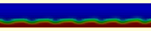} \\
		 BF1 & 
		 \includegraphics[width=0.3\linewidth]{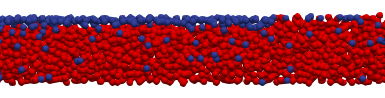} & \includegraphics[width=0.3\linewidth]{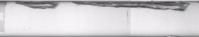} & \includegraphics[width=0.3\linewidth]{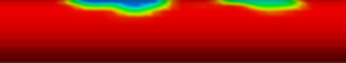} \\
		 SF & 
		 \includegraphics[width=0.3\linewidth]{Figs/intermittent_flow_patterns/slugFlow2.png} & \includegraphics[width=0.3\linewidth]{Figs/intermittent_flow_patterns/slugFlow2_exp.png} & \includegraphics[width=0.3\linewidth]{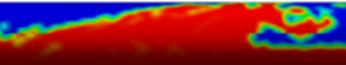} \\
		 BF2 & 
		 \includegraphics[width=0.3\linewidth]{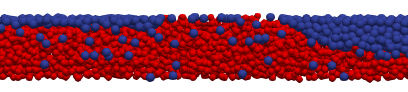} & \includegraphics[width=0.3\linewidth]{Figs/intermittent_flow_patterns/bubbleFlow_exp.png} & \includegraphics[width=0.3\linewidth]{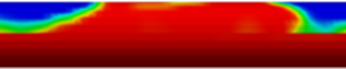} \\
	\end{tabular}
	\caption{Gas-liquid multi-phase flow in a horizontal pipe: Flow patterns obtained by the presented SPH method in 3-D
	under various 
	velocity conditions in comparison to the experimental and OpenFOAM simulations carried out by Wu et al.
	\cite{wu2021_frontiers}. The velocity condition and flow pattern description of the above flow regimes are
	summarized in Table \ref{table:pipeflowCharacteristics} denoted by the assigned ID.}
\label{fig:5pipeflowPatterns}
\end{figure}

The obtained results by the presented SPH method has been compared to the experimental observations of 
Wu et al.\cite{wu2021_frontiers} in Fig. \ref{fig:5pipeflowPatterns}.
As it is can be observed, the overall flow regime has been reproduced by both SPH and OpenFOAM 
in comparison with the experiments, while discrepancies are present in such a complex multi-phase flow.
It is also seen that some SPH particles from the gaseous phase have entered the liquid phase, which is due to the fact
that no special multi-phase interface sharpening technique has been employed in the presented SPH method
as explained in Ref. \cite{REZAVAND2020109092}.  
The present SPH simulations are in 3-D and with the large computational domain size in mind, we 
are dealing with a considerably large number of particle experiencing complex phenomena. For this reason,
the resolution of the SPH model is lower that of the OpenFOAM scenarios. 

\begin{figure}
	\centering
	\begin{subfigure}[b]{\textwidth}
		\centering
		\includegraphics[width=\textwidth]{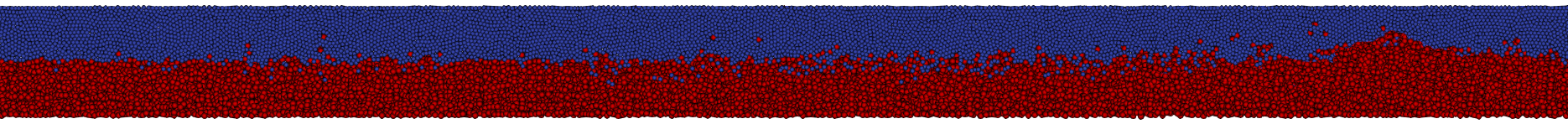}
		\caption{}
		\label{fig:slugFormationEntrance_a}
	\end{subfigure}
	\\
	\begin{subfigure}[b]{\textwidth}
		\centering
		\includegraphics[width=\textwidth]{Figs/slugFormationEntrance/slugEntranceExperiment.png}
		\caption{}
		\label{fig:slugFormationEntrance_b}
	\end{subfigure}
	\\
	\begin{subfigure}[b]{\textwidth}
		\centering
		\includegraphics[width=\textwidth]{Figs/slugFormationEntrance/slugEntranceOpenFOAM.png}
		\caption{}
		\label{fig:slugFormationEntrance_c}
	\end{subfigure}
	\caption{Gas-liquid multi-phase flow in a horizontal pipe: Slug formation process at the entrance of the pipe 
				obtained by (a) SPH, (b) experiments and (c) OpenFOAM. Both the experimental and the 
				OpenFOAM results are carried out by Wu et al.\cite{wu2021_frontiers}.}
	\label{fig:slugFormationEntrance}
\end{figure}

In a particular case, we have also considered the liquid slug formation process at the pipe entrance 
with the superficial velocity condition of $v_{sg}=2 ~\mathrm{m.s^{-1}}$ and $v_{sl}= 0.2~\mathrm{m.s^{-1}}$. 
From both the numerical simulations and the experimental observations 
presented in Fig. \ref{fig:slugFormationEntrance},
one can see that a series of high frequency waves with small amplitude are propagated at the gas-liquid interface.
After momentum transfer between gas and liquid in the slug flow regime, the wave frequency decreases while 
the amplitude increases. The pressure of gas gradually drops due to the Bernoulli effect and the pressure difference 
between gas and liquid dominates the tension and gravity forces at the gas-liquid interface.
Subsequently, the slug formation process initiates at the interface and the first liquid slug occurs at the entrance
of the pipe. 

Due to the prohibitive computational cost of 3-D simulations, both SPH and OpenFOAM simulations are carried out
in 2-D in Fig. \ref{fig:slugFormationEntrance}. The SPH simulation demonstrates a slightly lagged prediction of 
the slug formation, which is attributed to the turbulence model uncertainties that are not present in our methodology
influencing the momentum exchange between gas and liquid. As discussed in Section \ref{sec:introduction},
Ting et al. \cite{ting2005simulation} and Ghasemi et al. \cite{GhasemiV_2013} observed similar shortcomings 
when a 2-D SPH scheme was employed for a turbulent flow.
Apart from this observation, the overall slug formation process and the pre-slug high frequency waves 
have been well predicted by the presented method.

\section{Numerical results and discussion}\label{sec:ResultsandDiscussion}
In this section, we present the obtained numerical results for various aspects of the slug flow in detail
and discuss the insights gained by means of the presented 3-D computational framework based on SPH.
The slug flow investigated herein owns the same configuration as depicted in Fig. \ref{fig:slug_schematic}.

\begin{figure}
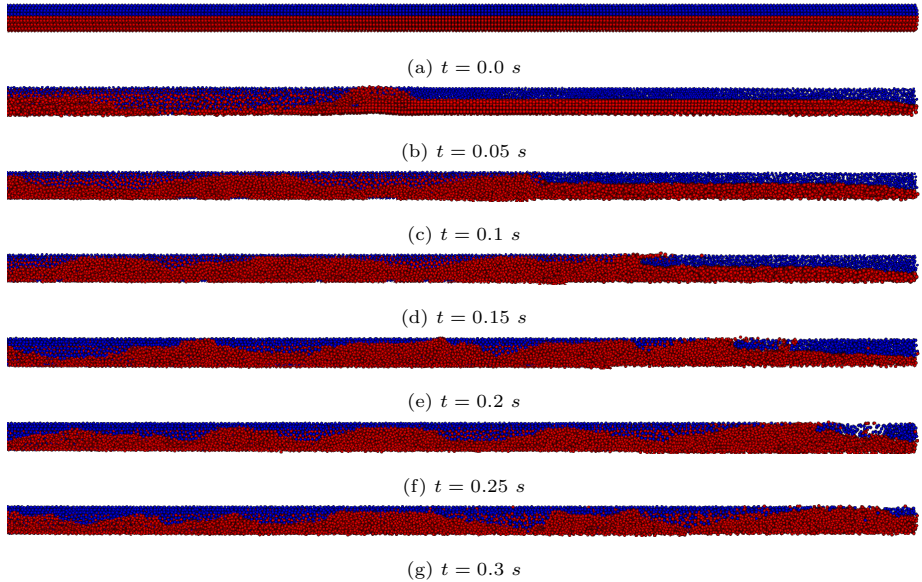

	\centering
	\begin{subfigure}[b]{\textwidth}
		\centering
		\includegraphics[width=\linewidth]{Figs/slugTimeSeries/timeSeries1.png}
		\caption{$t=0.0~s$}
		\label{fig:slugTimeSeries1}
	\end{subfigure}
	\hfill
	\begin{subfigure}[b]{\textwidth}
		\centering
		\includegraphics[width=\linewidth]{Figs/slugTimeSeries/timeSeries2.png}
		\caption{$t=0.05~s$}
		\label{fig:slugTimeSeries2}
	\end{subfigure}
	\hfill
	\begin{subfigure}[b]{\textwidth}
		\centering
		\includegraphics[width=\linewidth]{Figs/slugTimeSeries/timeSeries3.png}
		\caption{$t=0.1~s$}
		\label{fig:slugTimeSeries3}
	\end{subfigure}
	\hfill
	\begin{subfigure}[b]{\textwidth}
		\centering
		\includegraphics[width=\linewidth]{Figs/slugTimeSeries/timeSeries4.png}
		\caption{$t=0.15~s$}
		\label{fig:slugTimeSeries4}
	\end{subfigure}
	\hfill
	\begin{subfigure}[b]{\textwidth}
		\centering
		\includegraphics[width=\linewidth]{Figs/slugTimeSeries/timeSeries5.png}
		\caption{$t=0.2~s$}
		\label{fig:slugTimeSeries5}
	\end{subfigure}
	\hfill
	\begin{subfigure}[b]{\textwidth}
		\centering
		\includegraphics[width=\linewidth]{Figs/slugTimeSeries/timeSeries6.png}
		\caption{$t=0.25~s$}
		\label{fig:slugTimeSeries6}
	\end{subfigure}
	\hfill
	\begin{subfigure}[b]{\textwidth}
		\centering
		\includegraphics[width=\linewidth]{Figs/slugTimeSeries/timeSeries7.png}
		\caption{$t=0.3~s$}
		\label{fig:slugTimeSeries7}
	\end{subfigure}
	\caption{Gas-liquid multi-phase flow in a horizontal pipe: Time series of slug development 
		with the superficial velocity condition of $v_{sg}=3 ~\mathrm{m.s^{-1}}$ 
		and $v_{sl}= 1~\mathrm{m.s^{-1}}$.}
	\label{fig:slugTimeSeries}
\end{figure}

Fig. \ref{fig:slugTimeSeries} shows time series of slug development with the superficial velocity condition 
of $v_{sg}=3 ~\mathrm{m.s^{-1}}$ and $v_{sl}= 1~\mathrm{m.s^{-1}}$. As it is observed,
the first interaction of the high velocity gas with the slower liquid occurs at $t=0.05~s$ (Fig. \ref{fig:slugTimeSeries2}).
We will investigate this moment later in-depth considering velocity vectors of the both phases.
Meanwhile, one can note that from this point the momentum exchange between the phases accelerates and 
the first slug starts rising. At $t=0.15~s$ (Fig. \ref{fig:slugTimeSeries4}), the head of the slug has already jumped forward,
which later at $t=0.2~s$ (Fig. \ref{fig:slugTimeSeries5}) falls under the effect of the gravity acceleration. Throughout the
slug development, and in particular as the slug has passed by, low frequency waves with large amplitude are 
randomly generated due to the interfacial momentum transfer. 
It is also observed that slug roll-over happens, e.g., in Fig. \ref{fig:slugTimeSeries4}, which almost closes the channel.
Similar observations have been reported also previously in the literature 
(see e.g., \cite{vallee2008experimental,wu2021_frontiers,MOHMMED2021116611}).

\begin{figure}
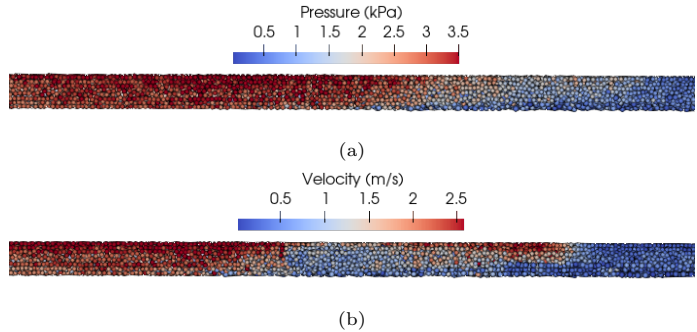

	\centering
	\begin{subfigure}[b]{0.75\textwidth}
		\centering
		\includegraphics[width=\textwidth]{Figs/pressure_distribution_t1.png}
		\caption{}
		\label{fig:slug_Pres_distribution}
	\end{subfigure}
	\begin{subfigure}[b]{\textwidth}
		\centering
		\includegraphics[width=0.75\textwidth]{Figs/velocity_distribution_t1.png}
		\caption{}
		\label{fig:slug_vel_distribution}
	\end{subfigure}
	\caption{Gas-liquid multi-phase flow in a horizontal pipe: Particle distribution together with
		(a) pressure and (b) velocity distribution.}
	\label{fig:slug_vel_Pres_distribution}
\end{figure}

Figs. \ref{fig:slug_Pres_distribution} and \ref{fig:slug_vel_distribution} show particle distribution together with
pressure and velocity distribution, respectively, obtained by the presented method.
It is noted that a quite smooth distribution of particles is achieved and pressure and velocity present no 
non-physical behavior, while differences in the
pressure gradient at the phase interface are notable. 
As the slugging phenomena is driven by the gas pressure,
a notable pressure gradient is also noted at the air-water interface.
Consequently, the upper part of the liquid particles at the top of the slug are violently driven, 
which causes the slug roll-over, shown in Fig. \ref{fig:slug_vel_distribution} as the high velocity
zone.

\begin{figure}
	\centering
	\frame{\includegraphics[width=0.75\linewidth]{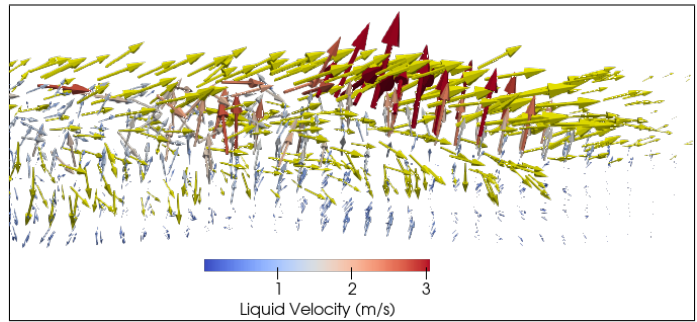}}
	\caption{Gas-liquid multi-phase flow in a horizontal pipe: Zoom-in view of the velocity vectors of both
		gas (in solid yellow) and liquid (scaled by their velocity magnitude) particles at $t=0.05~s$.}
	\label{fig:slug_liquidVelVectors}
\end{figure}

In order to further investigate the momentum exchange process causing the initiation of the first slug, 
Fig. \ref{fig:slug_liquidVelVectors} presents a zoom-in view of the first interaction between the high-velocity gas 
and the low-velocity liquid. In this figure, the solid yellow arrows represent the velocity vectors of the gas particles
and the arrows scaled with their velocity values are the liquid particles velocity vectors. The length of the 
vectors also represents the magnitude of velocity at that point. As it can be seen,
the interfacial interaction between gas and liquid gives rise to the velocity of the liquid (red arrows) and causes 
a fully non-linear velocity distribution. Within the lower levels of liquid, on the contrary, the velocity magnitude is 
at its minimum as it has also been observed in Fig. \ref{fig:slug_vel_distribution}.
As it can be also observed here, while the higher level air particles still have a relatively high velocity,
its lower level particles have already lost their high kinetic energy due to the momentum exchange with liquid. 
The complex flow structures at this time instant are also remarkable.

\begin{figure}
	\centering
	\includegraphics[width=0.75\linewidth]{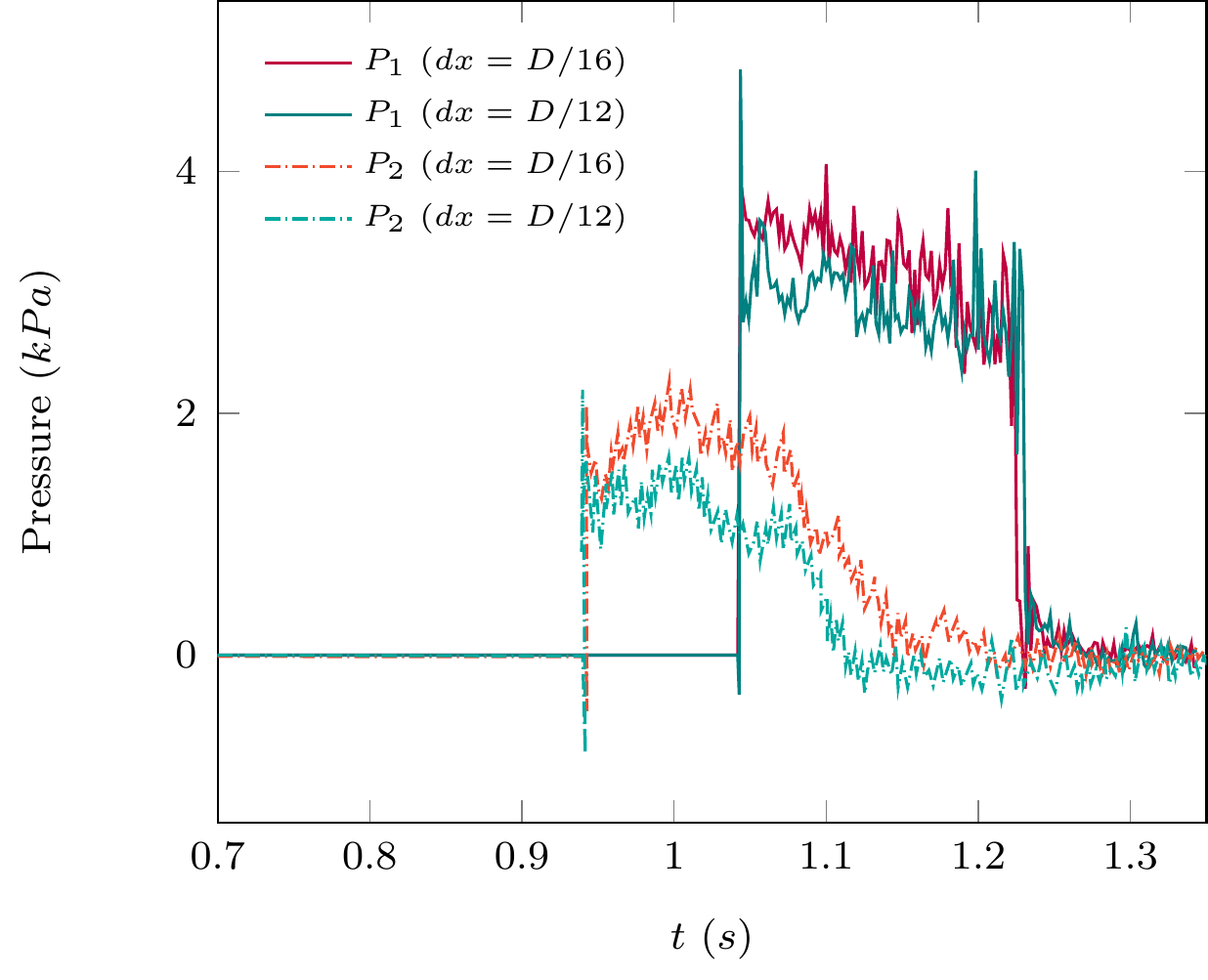}
	\caption{Gas-liquid multi-phase flow in a horizontal pipe: Transient pressure signals obtained at 
		$P_1$ and $P_2$ sensors described in Fig. \ref{fig:slug_schematic} for 
		the velocity condition of $v_{sg}=5 ~\mathrm{m.s^{-1}}$ 
		and $v_{sl}= 1~\mathrm{m.s^{-1}}$.}
	\label{fig:p1p2_signal}
\end{figure}

The above reflections from the initiation of the first slug can be followed-up in Fig. \ref{fig:p1p2_signal}.
The figure plots the transient pressure signals recorded at the two sensors shown in Fig. \ref{fig:slug_schematic}, 
$P_1$ and $P_2$.  As expected, an abrupt pressure increase has been recorded due to the rise of the first slug at both
sensors. The duration of the rise time depends on the position of sensors at the pipe and the fact that whether the slug
is developing or it is already developed. 
As the SPH formulations are in a weakly compressible framework, the pressure profile exhibits 
high frequency oscillations. It is worth noting that the measured pressure is obtained by 
averaging the values from particles within
a support radius of $2.6dx$ \cite{REZAVAND2020109092}.

\begin{figure}
	\centering
	\begin{subfigure}[b]{0.45\textwidth}
		\centering
		\includegraphics[width=\textwidth]{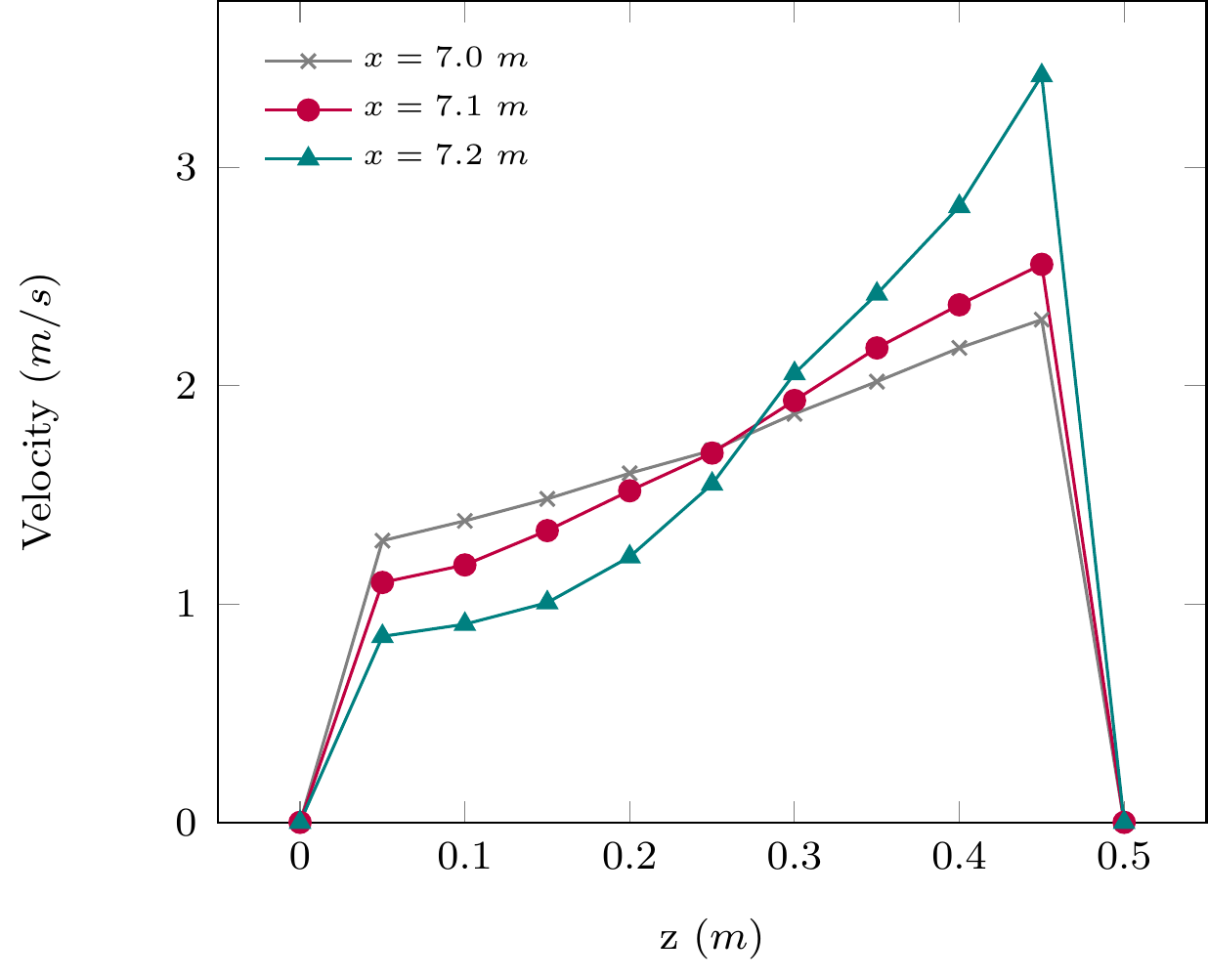}
		\caption{}
		\label{fig:slug_vel_developing}
	\end{subfigure}
	\hfill
	\begin{subfigure}[b]{0.45\textwidth}
		\centering
		\includegraphics[width=\textwidth]{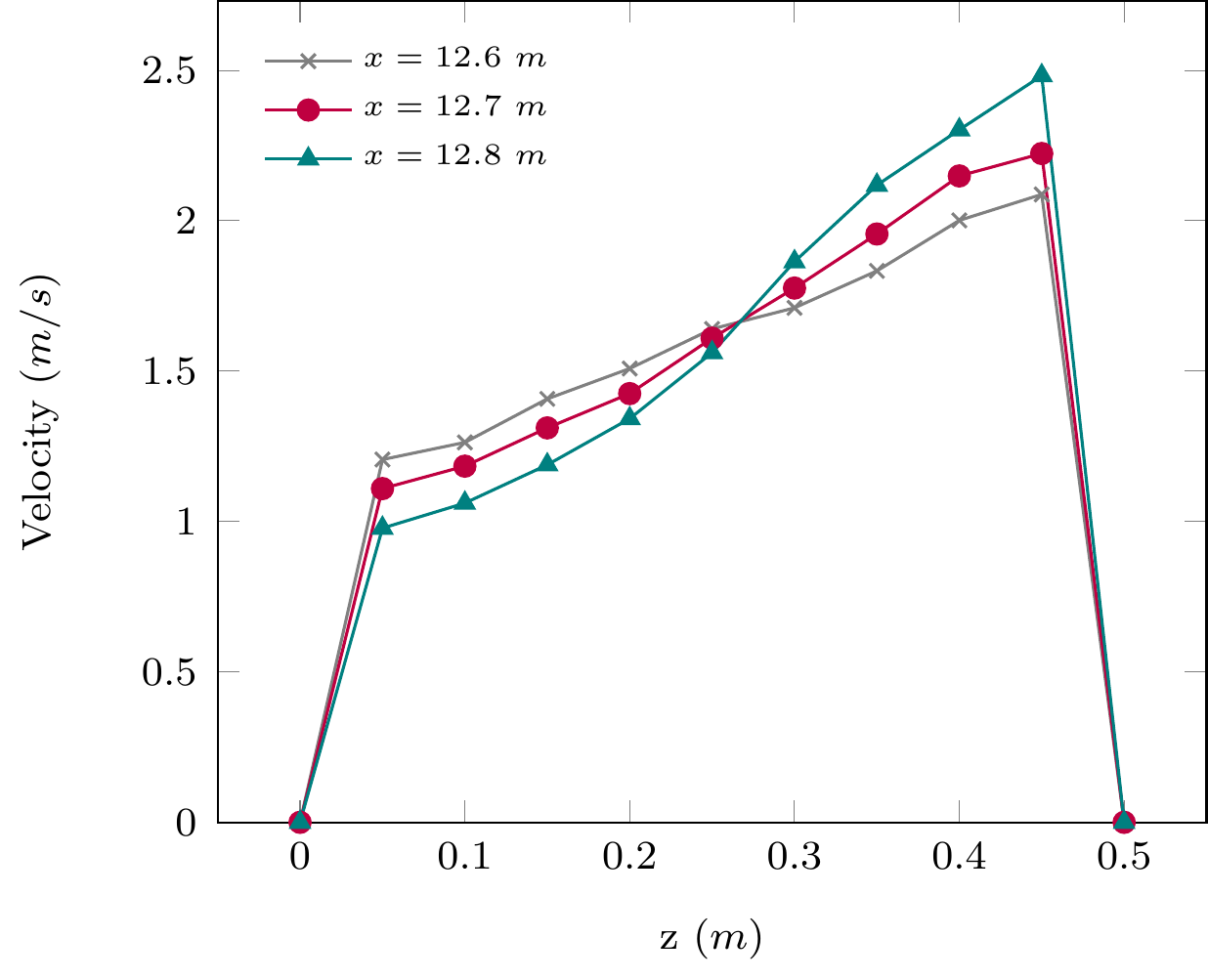}
		\caption{}
		\label{fig:slug_vel_developed}
	\end{subfigure}
	\caption{Gas-liquid multi-phase flow in a horizontal pipe: Velocity profile obtained for (a) developing and 
		(b) developed slug with the velocity condition of $v_{sg}=3 ~\mathrm{m.s^{-1}}$ 
		and $v_{sl}= 1~\mathrm{m.s^{-1}}$.}
	\label{fig:slug_vel_prifile}
\end{figure}

It is possible to evaluate the rate of development of the slugs by analyzing the cross-sectional velocity profile 
along the pipe. Fig. \ref{fig:slug_vel_prifile} shows the velocity profiles at two different sections of the pipe, 
each being measured by means of three sequential probes placed with small distances after each other with 
the superficial velocity condition of $v_{sg}=3 ~\mathrm{m.s^{-1}}$ 
and $v_{sl}= 1~\mathrm{m.s^{-1}}$.
It can be observed in Fig. \ref{fig:slug_vel_developing} that during the slug development 
the velocity of the liquid particles in higher vertical positions gradually increases along the slug length, 
whereas the velocity of the lower particles gradually drops. A similar velocity variation is noted between upper and lower
parts of the liquid slug, in particular at 
$x=7.2~m$, where the slug has already been initiated and its head demonstrates a high velocity as observed also in Figs.
\ref{fig:slug_vel_distribution} and \ref{fig:slug_liquidVelVectors}.
The velocity difference between upper and lower particles is mitigated in Fig. \ref{fig:slug_vel_developed}, where the slug
is fully developed.

\begin{figure}
	\centering
	\includegraphics[width=0.75\linewidth]{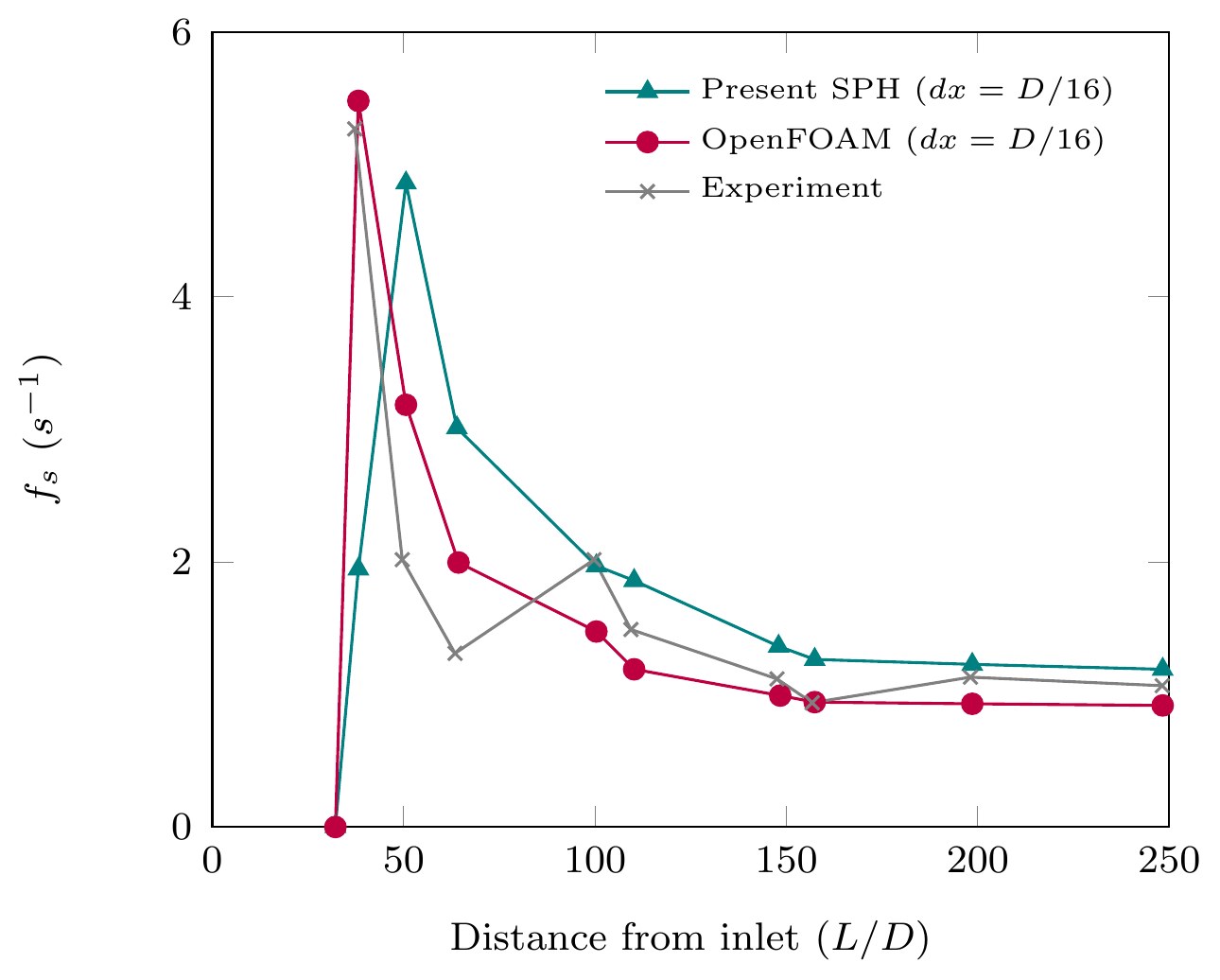}
	\caption{Gas-liquid multi-phase flow in a horizontal pipe: Slug frequency along the flow direction 
		obtained by the presented SPH method in comparison with experimental observations and OpenFOAM simulations
		\cite{wu2021_frontiers}.}
	\label{fig:frequency_profile}
\end{figure}

Another important characteristic quantity of liquid slugging is
the slug frequency, $f_s$, being defined as the inverse of the slug period, $T$, which is the time that 
a slug unit needs to pass through a measuring device
\begin{equation}
	f_s = \frac{1}{T}.
\end{equation}
Due to the prohibitive computational cost that impedes high resolution simulations, correctly detecting the slug head and tail 
to measure the slug period is not trivial. 
In this work we use the following formulation to measure an averaged slug frequency
\begin{equation}
\bar{f_s} = \frac{\sum_{i=1}^{N}{f_s}_{i}}{N}
\end{equation}
where $N$ is the number of times that the pressure sensors detect an abrupt pressure peak denoting 
a traveling slug. We placed ten pressure sensors along the pipe to detect the head and tail of a traveling slug by measuring 
the time between two different pressure peaks. Fig. \ref{fig:frequency_profile} plots the obtained slug frequencies along
the pipe in comparison with the numerical and experimental results of Wu et al. \cite{wu2021_frontiers} for
the velocity condition of $v_{sg}=3 ~\mathrm{m.s^{-1}}$ and $v_{sl}= 1~\mathrm{m.s^{-1}}$.
The slug frequency evidently decreases as the slug travels away from the pipe entrance where the dense and 
light fluids are injected into the pipe with maximum velocity. The overall profile of the slug frequency has been well predicted
with the SPH method, however, there are discrepancies between the SPH results 
and the experimental and numerical results of Wu et al. SPH demonstrates a delayed rise of frequency, which is consistence with 
the lagged slug initiation presented in Fig. \ref{fig:slugFormationEntrance_a}, although the computations related to Fig. \ref{fig:frequency_profile} have 
been carried out in 3-D. The delayed slug formation, as discussed in Section \ref{subsec:twophasePipe}, 
is attributed to the turbulence model uncertainties 
that influence the momentum exchange between gas and liquid, as well as the relatively coarse resolution dictated by the computational expenses. Finer resolution will demand a massively parallel framework as already presented by
Rezavand et al. \cite{REZAVAND2022108507} and will be the scope of future developments.
The SPH and OpenFOAM simulations used almost the same numerical resolution ($dx=D/16$), however, a higher rate of convergence is typically proven for
the finite volume method (FVM) than that of SPH presented in Fig. \ref{fig:hagenPoiseuille_vx_L2norm2} \cite{cai1990finite}.

\section{Conclusions}\label{sec:conclusions}
In this paper, we have numerically studied two-phase intermittent flows 
in a horizontal pipe and particularly focused on the slug flow regime. 
With the particle interactions being handled by a multi-phase SPH formulation based on a Riemann solver, 
the scheme is simple and the whole framework is not dependent on any empirical 
correlation for the slug flow. 

Whilst a few studies have performed 3-D simulations of slug flow with commercial packages or 
computationally expensive methods, e.g., DNS, the present study presents the first Lagrangian 3-D description of
the slug flows. The obtained results show that the mesh-free formulation makes it possible to 
explicitly realize the interfacial interactions between gas and liquid, with no need for a an interface tracking technique. 
The peculiarities of SPH in dealing with multi-phase phenomena facilitate the simulation of complex
two-phase pipe flows.

The proposed method is validated against existing experimental data for some aspects of the slug flow regime
and the results are verified in comparison with numerical and analytical solutions, as well, 
demonstrating reasonable agreements. 3-D description of the flow provides a more realistic 
representation of the studied real-world pipe flow.  

Albeit the present work focuses on benchmark test cases, 
the method is sufficiently generic to be employed in more complex industrial applications.
This will demand
a massively parallel numerical framework, which is the main scope of the future developments.
It is envisaged that the presented Lagrangian assets of SPH can improve the numerical analysis of slug flows
to achieve a better design of pipeline systems concerning safety measures.

\section*{Acknowledgments}
The authors gratefully acknowledge the financial support by Deutsche 
Forschungsgemeinschaft (DFG HU1572/10-1, DFG HU1527/12-4) for the present work.
The authors also would like to thank Zhaoting Wang and colleagues for providing their experimental and 
numerical results for validation purposes.

\section*{References}

\bibliography{bibliography}

\begin{thebibliography}{10}
\expandafter\ifx\csname url\endcsname\relax
  \def\url#1{\texttt{#1}}\fi
\expandafter\ifx\csname urlprefix\endcsname\relax\def\urlprefix{URL }\fi
\expandafter\ifx\csname href\endcsname\relax
  \def\href#1#2{#2} \def\path#1{#1}\fi

\bibitem{sausen2012slug}
A.~Sausen, P.~Sausen, M.~de~Campos, The slug flow problem in oil industry and
  pi level control, New technologies in the oil and gas industry (2012)
  103--118.

\bibitem{KirsnerProceedings}
W.~Kirsner, \href{https://doi.org/10.1115/PVP2005-71590}{{Condensation-Induced
  Water Hammer in District Steam Systems: Circumstances Resulting in
  Catastrophic Failures}} Volume 4: Fluid Structure Interaction (2005)
  875--881.
\newblock \href
  {http://arxiv.org/abs/https://asmedigitalcollection.asme.org/PVP/proceedings-pdf/PVP2005/41898/875/4560183/875\_1.pdf}
  {\path{arXiv:https://asmedigitalcollection.asme.org/PVP/proceedings-pdf/PVP2005/41898/875/4560183/875\_1.pdf}},
  \href {https://doi.org/10.1115/PVP2005-71590}
  {\path{doi:10.1115/PVP2005-71590}}.
\newline\urlprefix\url{https://doi.org/10.1115/PVP2005-71590}

\bibitem{bonizzi2003transient}
M.~Bonizzi, Transient one-dimensional modelling of multiphase slug flows, Ph.D.
  thesis, Imperial College London (University of London) (2003).

\bibitem{Fabre-annurev.fl.24.010192.000321}
J.~Fabre, A.~Line,
  \href{https://doi.org/10.1146/annurev.fl.24.010192.000321}{Modeling of
  two-phase slug flow}, Annual Review of Fluid Mechanics 24~(1) (1992) 21--46.
\newblock \href
  {http://arxiv.org/abs/https://doi.org/10.1146/annurev.fl.24.010192.000321}
  {\path{arXiv:https://doi.org/10.1146/annurev.fl.24.010192.000321}}, \href
  {https://doi.org/10.1146/annurev.fl.24.010192.000321}
  {\path{doi:10.1146/annurev.fl.24.010192.000321}}.
\newline\urlprefix\url{https://doi.org/10.1146/annurev.fl.24.010192.000321}

\bibitem{ujang2003studies}
P.~M. Ujang, Studies of slug initiation and development in two-phase gas-liquid
  pipeline flow, Ph.D. thesis, Imperial College London (University of London)
  (2003).

\bibitem{mandhane1974flow}
J.~Mandhane, G.~Gregory, K.~Aziz, A flow pattern map for gas—liquid flow in
  horizontal pipes, International journal of multiphase flow 1~(4) (1974)
  537--553.

\bibitem{mishima1980theoretical}
K.~Mishima, M.~Ishii, \href{https://doi.org/10.1115/1.3240720}{{Theoretical
  Prediction of Onset of Horizontal Slug Flow}}, Journal of Fluids Engineering
  102~(4) (1980) 441--445.
\newblock \href
  {http://arxiv.org/abs/https://asmedigitalcollection.asme.org/fluidsengineering/article-pdf/102/4/441/5531529/441\_1.pdf}
  {\path{arXiv:https://asmedigitalcollection.asme.org/fluidsengineering/article-pdf/102/4/441/5531529/441\_1.pdf}},
  \href {https://doi.org/10.1115/1.3240720} {\path{doi:10.1115/1.3240720}}.
\newline\urlprefix\url{https://doi.org/10.1115/1.3240720}

\bibitem{barnea1987unified}
D.~Barnea, A unified model for predicting flow-pattern transitions for the
  whole range of pipe inclinations, International journal of multiphase flow
  13~(1) (1987) 1--12.

\bibitem{taitel1976model}
Y.~Taitel, A.~E. Dukler, A model for predicting flow regime transitions in
  horizontal and near horizontal gas-liquid flow, AIChE journal 22~(1) (1976)
  47--55.

\bibitem{athlete3-2}
G.~Lerchl, H.~Austregesilo, A.~Langenfeld, P.~J. Schoeffel, D.~von~der Cron,
  F.~Weyermann, Athlet 3.2 - user’s manual, Tech. rep., GRS, GRS‐P‐1/
  Vol. 1 Rev.8 (2019).

\bibitem{chung2010mars}
B.~D. Chung, K.~D. Kim, S.~W. Bae, J.~J. Jeong, S.~W. Lee, M.~K. Hwang,
  C.~Yoon, Mars code manual volume i: code structure, system models, and
  solution methods, Tech. rep., Korea Atomic Energy Research Institute (2010).

\bibitem{bajorek2008trace}
S.~Bajorek, et~al., Trace v5. 0 theory manual, Field Equations, Solution
  Methods and Physical Models, US Nuclear Regulatory Commission (2008).

\bibitem{BloemelingNeuhausSchaffrath2013}
F.~Bloemeling, T.~Neuhaus, A.~Schaffrath,
  \href{https://doi.org/10.3139/124.110308}{1d models for condensation induced
  water hammer in pipelines}, Kerntechnik 78~(1) (2013) 31--34 [cited
  2023-02-19].
\newblock \href {https://doi.org/doi:10.3139/124.110308}
  {\path{doi:doi:10.3139/124.110308}}.
\newline\urlprefix\url{https://doi.org/10.3139/124.110308}

\bibitem{LEE2022112066}
J.~Lee, M.~Junk, T.~Skorek, P.~{Josef Schöffel},
  \href{https://www.sciencedirect.com/science/article/pii/S0029549322004174}{Improvement
  of entrainment model for horizontal flow in athlet and application to
  mantilla experiment and tptf}, Nuclear Engineering and Design 400 (2022)
  112066.
\newblock \href
  {https://doi.org/https://doi.org/10.1016/j.nucengdes.2022.112066}
  {\path{doi:https://doi.org/10.1016/j.nucengdes.2022.112066}}.
\newline\urlprefix\url{https://www.sciencedirect.com/science/article/pii/S0029549322004174}

\bibitem{ishii1975thermo}
M.~Ishii, Thermo-fluid dynamic theory of two-phase flow, NASA Sti/recon
  Technical Report A 75 (1975) 29657.

\bibitem{lu2015experimental}
M.~Lu, Experimental and computational study of two-phase slug flow, Ph.D.
  thesis, Imperial College London UK (2015).

\bibitem{korzilius2017modeling}
S.~P. Korzilius, A.~S. Tijsseling, Z.~Bozku{\c{s}}, M.~J. Anthonissen, W.~H.
  Schilders, Modeling liquid slugs accelerating in inclined conduits, Journal
  of Pressure Vessel Technology 139~(6) (2017).

\bibitem{vallee2008experimental}
C.~Vall{\'e}e, T.~H{\"o}hne, H.-M. Prasser, T.~S{\"u}hnel, Experimental
  investigation and cfd simulation of horizontal stratified two-phase flow
  phenomena, Nuclear Engineering and Design 238~(3) (2008) 637--646.

\bibitem{ABDULKADIR2016147}
M.~Abdulkadir, V.~Hernandez-Perez, I.~Lowndes, B.~Azzopardi, E.~Sam-Mbomah,
  \href{https://www.sciencedirect.com/science/article/pii/S0009250916304948}{Experimental
  study of the hydrodynamic behaviour of slug flow in a horizontal pipe},
  Chemical Engineering Science 156 (2016) 147--161.
\newblock \href {https://doi.org/https://doi.org/10.1016/j.ces.2016.09.015}
  {\path{doi:https://doi.org/10.1016/j.ces.2016.09.015}}.
\newline\urlprefix\url{https://www.sciencedirect.com/science/article/pii/S0009250916304948}

\bibitem{wu2021_frontiers}
X.~Wu, Z.~Wang, M.~Dong, Q.~Ge, L.~Dong,
  \href{https://www.frontiersin.org/articles/10.3389/fenrg.2021.762471}{Simulation
  study on the development process and phase interface structure of gas-liquid
  slug flow in a horizontal pipe}, Frontiers in Energy Research 9 (2021).
\newblock \href {https://doi.org/10.3389/fenrg.2021.762471}
  {\path{doi:10.3389/fenrg.2021.762471}}.
\newline\urlprefix\url{https://www.frontiersin.org/articles/10.3389/fenrg.2021.762471}

\bibitem{ramdin2012computational}
M.~Ramdin, R.~Henkes, Computational fluid dynamics modeling of benjamin and
  taylor bubbles in two-phase flow in pipes, Journal of fluids engineering
  134~(4) (2012).

\bibitem{TAHA20041181}
T.~Taha, Z.~Cui,
  \href{https://www.sciencedirect.com/science/article/pii/S0009250903005979}{Hydrodynamics
  of slug flow inside capillaries}, Chemical Engineering Science 59~(6) (2004)
  1181--1190.
\newblock \href {https://doi.org/https://doi.org/10.1016/j.ces.2003.10.025}
  {\path{doi:https://doi.org/10.1016/j.ces.2003.10.025}}.
\newline\urlprefix\url{https://www.sciencedirect.com/science/article/pii/S0009250903005979}

\bibitem{al2016numerical}
Z.~I. Al-Hashimy, H.~H. Al-Kayiem, R.~W. Time, Z.~K. Kadhim, Numerical
  characterisation of slug flow in horizontal air/water pipe flow,
  International Journal of Computational Methods and Experimental Measurements
  4~(2) (2016) 114--130.

\bibitem{FUKAGATA200772}
K.~Fukagata, N.~Kasagi, P.~Ua-arayaporn, T.~Himeno,
  \href{https://www.sciencedirect.com/science/article/pii/S0142727X06001354}{Numerical
  simulation of gas–liquid two-phase flow and convective heat transfer in a
  micro tube}, International Journal of Heat and Fluid Flow 28~(1) (2007)
  72--82, the International Conference on Heat Transfer and Fluid Flow in
  Microscale (HTFFM-05).
\newblock \href
  {https://doi.org/https://doi.org/10.1016/j.ijheatfluidflow.2006.04.010}
  {\path{doi:https://doi.org/10.1016/j.ijheatfluidflow.2006.04.010}}.
\newline\urlprefix\url{https://www.sciencedirect.com/science/article/pii/S0142727X06001354}

\bibitem{YU20077172}
Z.~Yu, O.~Hemminger, L.-S. Fan,
  \href{https://www.sciencedirect.com/science/article/pii/S0009250907006884}{Experiment
  and lattice boltzmann simulation of two-phase gas–liquid flows in
  microchannels}, Chemical Engineering Science 62~(24) (2007) 7172--7183, 8th
  International Conference on Gas-Liquid and Gas-Liquid-Solid Reactor
  Engineering.
\newblock \href {https://doi.org/https://doi.org/10.1016/j.ces.2007.08.075}
  {\path{doi:https://doi.org/10.1016/j.ces.2007.08.075}}.
\newline\urlprefix\url{https://www.sciencedirect.com/science/article/pii/S0009250907006884}

\bibitem{xie_zheng_2017}
F.~Xie, X.~Zheng, M.~S. Triantafyllou, Y.~Constantinides, Y.~Zheng,
  G.~Em~Karniadakis, Direct numerical simulations of two-phase flow in an
  inclined pipe, Journal of Fluid Mechanics 825 (2017) 189–207.
\newblock \href {https://doi.org/10.1017/jfm.2017.417}
  {\path{doi:10.1017/jfm.2017.417}}.

\bibitem{rezavand2018isph}
M.~Rezavand, M.~Taeibi-Rahni, W.~Rauch, {An ISPH scheme for numerical
  simulation of multiphase flows with complex interfaces and high density
  ratios}, Comput. Math. Appl. 75~(8) (2018) 2658--2677.
\newblock \href {https://doi.org/10.1016/j.camwa.2017.12.034}
  {\path{doi:10.1016/j.camwa.2017.12.034}}.

\bibitem{Lou-Khayyer.2022.jc873}
X.~Su, M.~Luo, X.~Zhao, A.~Khayyer,
  \href{https://doi.org/10.17736/ijope.2022.jc873}{{Oil Spill Spreading
  Simulation Based on an Enhanced Multi-phase Consistent Particle Method}},
  International Journal of Offshore and Polar Engineering 32~(04) (2022)
  377--385.
\newblock \href
  {http://arxiv.org/abs/https://onepetro.org/IJOPE/article-pdf/32/04/377/3055546/isope-22-32-4-377.pdf}
  {\path{arXiv:https://onepetro.org/IJOPE/article-pdf/32/04/377/3055546/isope-22-32-4-377.pdf}},
  \href {https://doi.org/10.17736/ijope.2022.jc873}
  {\path{doi:10.17736/ijope.2022.jc873}}.
\newline\urlprefix\url{https://doi.org/10.17736/ijope.2022.jc873}

\bibitem{PATINONARINO2023104355}
E.~A. Patiño-Nariño, A.~F. Galvis, R.~Pavanello, M.~R. Gongora-Rubio,
  \href{https://www.sciencedirect.com/science/article/pii/S0301932222003147}{Modeling
  of co-axial bubbles coalescence under moderate reynolds regimes: A bi-phase
  sph approach}, International Journal of Multiphase Flow 162 (2023) 104355.
\newblock \href
  {https://doi.org/https://doi.org/10.1016/j.ijmultiphaseflow.2022.104355}
  {\path{doi:https://doi.org/10.1016/j.ijmultiphaseflow.2022.104355}}.
\newline\urlprefix\url{https://www.sciencedirect.com/science/article/pii/S0301932222003147}

\bibitem{REZAVAND2020109092}
M.~Rezavand, C.~Zhang, X.~Hu,
  \href{https://www.sciencedirect.com/science/article/pii/S0021999119307971}{A
  weakly compressible sph method for violent multi-phase flows with high
  density ratio}, Journal of Computational Physics 402 (2020) 109092.
\newblock \href {https://doi.org/https://doi.org/10.1016/j.jcp.2019.109092}
  {\path{doi:https://doi.org/10.1016/j.jcp.2019.109092}}.
\newline\urlprefix\url{https://www.sciencedirect.com/science/article/pii/S0021999119307971}

\bibitem{ZHANG_sphinxsys}
C.~Zhang, M.~Rezavand, Y.~Zhu, Y.~Yu, D.~Wu, W.~Zhang, J.~Wang, X.~Hu,
  \href{https://www.sciencedirect.com/science/article/pii/S0010465521001788}{Sphinxsys:
  An open-source multi-physics and multi-resolution library based on smoothed
  particle hydrodynamics}, Computer Physics Communications 267 (2021) 108066.
\newblock \href {https://doi.org/https://doi.org/10.1016/j.cpc.2021.108066}
  {\path{doi:https://doi.org/10.1016/j.cpc.2021.108066}}.
\newline\urlprefix\url{https://www.sciencedirect.com/science/article/pii/S0010465521001788}

\bibitem{ZHANG2021_cardiac}
C.~Zhang, J.~Wang, M.~Rezavand, D.~Wu, X.~Hu,
  \href{https://www.sciencedirect.com/science/article/pii/S0045782521001845}{An
  integrative smoothed particle hydrodynamics method for modeling cardiac
  function}, Computer Methods in Applied Mechanics and Engineering 381 (2021)
  113847.
\newblock \href {https://doi.org/https://doi.org/10.1016/j.cma.2021.113847}
  {\path{doi:https://doi.org/10.1016/j.cma.2021.113847}}.
\newline\urlprefix\url{https://www.sciencedirect.com/science/article/pii/S0045782521001845}

\bibitem{KHAYYER202384}
A.~Khayyer, Y.~Shimizu, T.~Gotoh, H.~Gotoh,
  \href{https://www.sciencedirect.com/science/article/pii/S0307904X22005091}{Enhanced
  resolution of the continuity equation in explicit weakly compressible sph
  simulations of incompressible free‐surface fluid flows}, Applied
  Mathematical Modelling 116 (2023) 84--121.
\newblock \href {https://doi.org/https://doi.org/10.1016/j.apm.2022.10.037}
  {\path{doi:https://doi.org/10.1016/j.apm.2022.10.037}}.
\newline\urlprefix\url{https://www.sciencedirect.com/science/article/pii/S0307904X22005091}

\bibitem{DOUILLETGRELLIER2018101}
T.~Douillet-Grellier, F.~{De Vuyst}, H.~Calandra, P.~Ricoux,
  \href{https://www.sciencedirect.com/science/article/pii/S0045793018307199}{Simulations
  of intermittent two-phase flows in pipes using smoothed particle
  hydrodynamics}, Computers \& Fluids 177 (2018) 101--122.
\newblock \href
  {https://doi.org/https://doi.org/10.1016/j.compfluid.2018.10.004}
  {\path{doi:https://doi.org/10.1016/j.compfluid.2018.10.004}}.
\newline\urlprefix\url{https://www.sciencedirect.com/science/article/pii/S0045793018307199}

\bibitem{douillet2019comparison}
T.~Douillet-Grellier, S.~Leclaire, D.~Vidal, F.~Bertrand, F.~De~Vuyst,
  Comparison of multiphase sph and lbm approaches for the simulation of
  intermittent flows, Computational Particle Mechanics 6 (2019) 695--720.

\bibitem{GhasemiV_2013}
A.~{Ghasemi V.}, B.~Firoozabadi, M.~Mahdinia, {2D numerical simulation of
  density currents using the SPH projection method}, Eur. J. Mech. B. Fluids 38
  (2013) 38--46.
\newblock \href {https://doi.org/10.1016/j.euromechflu.2012.10.004}
  {\path{doi:10.1016/j.euromechflu.2012.10.004}}.

\bibitem{cleary1993boundary}
P.~W. Cleary, J.~J. Monaghan, Boundary interactions and transition to
  turbulence for standard cfd problems using sph, Proc. of the 6th
  International Computational Techniques and Applications (1993) 157--165.

\bibitem{ting2005simulation}
T.~S. Ting, M.~Prakash, P.~Cleary, M.~Thompson, Simulation of high reynolds
  number flow over a backward facing step using sph, ANZIAM Journal 47 (2005)
  C292--C309.

\bibitem{vacondio2021grand}
R.~Vacondio, C.~Altomare, M.~De~Leffe, X.~Hu, D.~Le~Touz{\'e}, S.~Lind, J.-C.
  Marongiu, S.~Marrone, B.~D. Rogers, A.~Souto-Iglesias, Grand challenges for
  smoothed particle hydrodynamics numerical schemes, Computational Particle
  Mechanics 8 (2021) 575--588.

\bibitem{MOHMMED2021116611}
A.~O. Mohmmed, H.~H. Al-Kayiem, A.~Osman,
  \href{https://www.sciencedirect.com/science/article/pii/S0009250921001767}{Investigations
  on the slug two-phase flow in horizontal pipes: Past, presents, and future
  directives}, Chemical Engineering Science 238 (2021) 116611.
\newblock \href {https://doi.org/https://doi.org/10.1016/j.ces.2021.116611}
  {\path{doi:https://doi.org/10.1016/j.ces.2021.116611}}.
\newline\urlprefix\url{https://www.sciencedirect.com/science/article/pii/S0009250921001767}

\bibitem{zhang2021sphinxsys}
C.~Zhang, M.~Rezavand, Y.~Zhu, Y.~Yu, D.~Wu, W.~Zhang, J.~Wang, X.~Hu,
  Sphinxsys: an open-source multi-physics and multi-resolution library based on
  smoothed particle hydrodynamics, Computer Physics Communications (2021)
  108066\href {https://doi.org/https://doi.org/10.1016/j.cpc.2021.108066}
  {\path{doi:https://doi.org/10.1016/j.cpc.2021.108066}}.

\bibitem{REZAVAND2022108507}
M.~Rezavand, C.~Zhang, X.~Hu,
  \href{https://www.sciencedirect.com/science/article/pii/S0010465522002260}{Generalized
  and efficient wall boundary condition treatment in gpu-accelerated smoothed
  particle hydrodynamics}, Computer Physics Communications 281 (2022) 108507.
\newblock \href {https://doi.org/https://doi.org/10.1016/j.cpc.2022.108507}
  {\path{doi:https://doi.org/10.1016/j.cpc.2022.108507}}.
\newline\urlprefix\url{https://www.sciencedirect.com/science/article/pii/S0010465522002260}

\bibitem{Monaghan_2012_Annual_Rew}
J.~J. Monaghan, {Smoothed Particle Hydrodynamics and Its Diverse Applications},
  Annu. Rev. Fluid Mech. 44~(1) (2012) 323--346.
\newblock \href {https://doi.org/10.1146/annurev-fluid-120710-101220}
  {\path{doi:10.1146/annurev-fluid-120710-101220}}.

\bibitem{vila1999particle}
J.~Vila, On particle weighted methods and smooth particle hydrodynamics, Math.
  Models Methods Appl. Sci. 9~(02) (1999) 161--209.
\newblock \href {https://doi.org/https://doi.org/10.1142/S0218202599000117}
  {\path{doi:https://doi.org/10.1142/S0218202599000117}}.

\bibitem{zhang2017weakly}
C.~Zhang, X.~Hu, N.~A. Adams, {A weakly compressible SPH method based on a
  low-dissipation Riemann solver}, J. Comput. Phys. 335 (2017) 605--620.
\newblock \href {https://doi.org/https://doi.org/10.1016/j.jcp.2017.01.027}
  {\path{doi:https://doi.org/10.1016/j.jcp.2017.01.027}}.

\bibitem{hu2004interface}
X.~Y. Hu, B.~C. Khoo, An interface interaction method for compressible
  multifluids, J. Comput. Phys. 198~(1) (2004) 35--64.
\newblock \href {https://doi.org/10.1016/j.jcp.2003.12.018}
  {\path{doi:10.1016/j.jcp.2003.12.018}}.

\bibitem{ADAMI2012wall}
S.~Adami, X.~Hu, N.~Adams, A generalized wall boundary condition for smoothed
  particle hydrodynamics, J. Comput. Phys. 231~(21) (2012) 7057 -- 7075.
\newblock \href {https://doi.org/10.1016/j.jcp.2012.05.005}
  {\path{doi:10.1016/j.jcp.2012.05.005}}.

\bibitem{Shuoguo_arXiv}
S.~Zhang, W.~Zhang, C.~Zhang, X.~Hu, \href{https://arxiv.org/abs/2206.06875}{A
  lagrangian free-stream boundary condition for weakly compressible smoothed
  particle hydrodynamics} (2022).
\newblock \href {https://doi.org/10.48550/ARXIV.2206.06875}
  {\path{doi:10.48550/ARXIV.2206.06875}}.
\newline\urlprefix\url{https://arxiv.org/abs/2206.06875}

\bibitem{Lastiwka2009permeable}
M.~Lastiwka, M.~Basa, N.~J. Quinlan,
  \href{https://onlinelibrary.wiley.com/doi/abs/10.1002/fld.1971}{Permeable and
  non-reflecting boundary conditions in sph}, International Journal for
  Numerical Methods in Fluids 61~(7) (2009) 709--724.
\newblock \href
  {http://arxiv.org/abs/https://onlinelibrary.wiley.com/doi/pdf/10.1002/fld.1971}
  {\path{arXiv:https://onlinelibrary.wiley.com/doi/pdf/10.1002/fld.1971}},
  \href {https://doi.org/https://doi.org/10.1002/fld.1971}
  {\path{doi:https://doi.org/10.1002/fld.1971}}.
\newline\urlprefix\url{https://onlinelibrary.wiley.com/doi/abs/10.1002/fld.1971}

\bibitem{monaghan2005}
J.~J. Monaghan, \href{http://stacks.iop.org/0034-4885/68/i=8/a=R01}{Smoothed
  particle hydrodynamics}, Rep. Prog. Phys. 68~(8) (2005) 1703.
\newline\urlprefix\url{http://stacks.iop.org/0034-4885/68/i=8/a=R01}

\bibitem{adami2013transport}
S.~Adami, X.~Hu, N.~A. Adams, A transport-velocity formulation for smoothed
  particle hydrodynamics, J. Comput. Phys. 241 (2013) 292--307.
\newblock \href {https://doi.org/10.1016/j.jcp.2013.01.043}
  {\path{doi:10.1016/j.jcp.2013.01.043}}.

\bibitem{ZHANG2020109135}
C.~Zhang, M.~Rezavand, X.~Hu,
  \href{https://www.sciencedirect.com/science/article/pii/S002199911930840X}{Dual-criteria
  time stepping for weakly compressible smoothed particle hydrodynamics},
  Journal of Computational Physics 404 (2020) 109135.
\newblock \href {https://doi.org/https://doi.org/10.1016/j.jcp.2019.109135}
  {\path{doi:https://doi.org/10.1016/j.jcp.2019.109135}}.
\newline\urlprefix\url{https://www.sciencedirect.com/science/article/pii/S002199911930840X}

\bibitem{dongWu2022}
D.~Wu, C.~Zhang, X.~Tang, X.~Hu, \href{https://arxiv.org/abs/2212.00753}{An
  hourglass-free formulation for total lagrangian smoothed particle
  hydrodynamics} (2022).
\newblock \href {https://doi.org/10.48550/ARXIV.2212.00753}
  {\path{doi:10.48550/ARXIV.2212.00753}}.
\newline\urlprefix\url{https://arxiv.org/abs/2212.00753}

\bibitem{ZHANG2021_multiFSI}
C.~Zhang, M.~Rezavand, X.~Hu,
  \href{https://www.sciencedirect.com/science/article/pii/S0021999120308020}{A
  multi-resolution sph method for fluid-structure interactions}, Journal of
  Computational Physics 429 (2021) 110028.
\newblock \href {https://doi.org/https://doi.org/10.1016/j.jcp.2020.110028}
  {\path{doi:https://doi.org/10.1016/j.jcp.2020.110028}}.
\newline\urlprefix\url{https://www.sciencedirect.com/science/article/pii/S0021999120308020}

\bibitem{REN2023113110}
Y.~Ren, A.~Khayyer, P.~Lin, X.~Hu,
  \href{https://www.sciencedirect.com/science/article/pii/S0029801822023939}{Numerical
  modeling of sloshing flow interaction with an elastic baffle using
  sphinxsys}, Ocean Engineering 267 (2023) 113110.
\newblock \href
  {https://doi.org/https://doi.org/10.1016/j.oceaneng.2022.113110}
  {\path{doi:https://doi.org/10.1016/j.oceaneng.2022.113110}}.
\newline\urlprefix\url{https://www.sciencedirect.com/science/article/pii/S0029801822023939}

\bibitem{ZHANG20221_damping}
C.~Zhang, Y.~Zhu, Y.~Yu, D.~Wu, M.~Rezavand, S.~Shao, X.~Hu,
  \href{https://www.sciencedirect.com/science/article/pii/S0955799722001783}{An
  artificial damping method for total lagrangian sph method with application in
  biomechanics}, Engineering Analysis with Boundary Elements 143 (2022) 1--13.
\newblock \href
  {https://doi.org/https://doi.org/10.1016/j.enganabound.2022.05.022}
  {\path{doi:https://doi.org/10.1016/j.enganabound.2022.05.022}}.
\newline\urlprefix\url{https://www.sciencedirect.com/science/article/pii/S0955799722001783}

\bibitem{MORRIS1997214}
J.~P. Morris, P.~J. Fox, Y.~Zhu,
  \href{https://www.sciencedirect.com/science/article/pii/S0021999197957764}{Modeling
  low reynolds number incompressible flows using sph}, Journal of Computational
  Physics 136~(1) (1997) 214--226.
\newblock \href {https://doi.org/https://doi.org/10.1006/jcph.1997.5776}
  {\path{doi:https://doi.org/10.1006/jcph.1997.5776}}.
\newline\urlprefix\url{https://www.sciencedirect.com/science/article/pii/S0021999197957764}

\bibitem{Bird_2002_transport}
R.~B. Bird, W.~E. Stewart, E.~N. Lightfoot, {Transport phenomena}, New York:
  John Wiley \& Sons, Inc, 2002.

\bibitem{Fatehi_2011_error}
R.~Fatehi, M.~Manzari, {Error estimation in smoothed particle hydrodynamics and
  a new scheme for second derivatives}, Computers \& Mathematics with
  Applications 61~(2) (2011) 482--498.
\newblock \href {https://doi.org/10.1016/j.camwa.2010.11.028}
  {\path{doi:10.1016/j.camwa.2010.11.028}}.

\bibitem{cai1990finite}
Z.~Cai, On the finite volume element method, Numerische Mathematik 58~(1)
  (1990) 713--735.

\end{thebibliography}

\end{document}